\documentclass[fleqn,usenatbib]{mnras}
\usepackage{lineno} 

\usepackage[T1]{fontenc}
\usepackage{ae,aecompl}

\usepackage{xfrac}
\usepackage{float}
\usepackage{physics}
\usepackage{amssymb}
\usepackage{amsmath}
\usepackage{savesym}
\savesymbol{tablenum}
\usepackage{rotating}
\usepackage[section]{placeins}
\usepackage[noabbrev]{cleveref}
\usepackage{subfigure}
\usepackage{gensymb}
\usepackage{siunitx}
\usepackage{changepage}

\newcommand{\HI}{{H}{I}}

\newcommand{\kms}{km s$^{-1}$}
\newcommand{\dunit}{cm \textsuperscript{-3}}
\restoresymbol{SIX}{tablenum}
\newcommand{\tus}[1]{$_{\text{2}}$}
\newcommand{\hh}{$H_2$}

\newcommand{\los}{$V_{los}$}

\title[The kinematics of multi-phase Splash Bridges]{The generation of a multi-phase medium in ``Splash'' bridge systems: Towards an understanding of star formation suppression in turbulent galaxy systems}

\author[Yeager, Struck and Appleton]{
Travis R. Yeager,$^{1}$\thanks{E-mail: yeagerastro@gmail.com}
Curtis Struck,$^{2}$ and Phil Appleton,$^{3}$
\\
$^{1}$Lawrence Livermore National Laboratory, 7000 East Ave, Livermore, CA, 94550, USA\\
$^{2}$Iowa State University, Physics Hall, 2323 Osborn Dr \#12, Ames, IA, 50011, USA\\
$^{3}$Caltech/IPAC, MC 314-6, 1200 E. California Blvd, Pasadena, CA, 91125, USA\\
}

\date{Accepted 2024 September 16. Received 2024 July 29; in original form 2024 March 05.}

\pubyear{2023}

\begin{document}
\label{firstpage}
\pagerange{\pageref{firstpage}--\pageref{lastpage}}
\maketitle

\begin{abstract}
Cloud-cloud collisions in splash bridges produced in gas-rich disk galaxy collisions offer a brief but interesting environment to study the effects of shocks and turbulence on star formation rates in the diffuse IGM, far from the significant feedback effects of massive star formation and AGN. Expanding on our earlier work, we describe simulated collisions between counter-rotating disk galaxies of relatively similar mass, focusing on the thermal and kinematic effects of relative inclination and disk offset at the closest approach. This includes essential heating and cooling signatures, which go some way towards explaining the luminous power in H$_2$ and \textnormal{[CII]} emission in the Taffy bridge, as well as providing a partial explanation of the turbulent nature of the recently observed compact CO-emitting clouds observed in Taffy by ALMA. The models show counter-rotating disk collisions result in swirling, shearing kinematics for the gas in much of the post-collision bridge. Gas with little specific angular momentum due to collisions between counter-rotating streams accumulates near the center of mass. The disturbances and mixing in the bridge drive continuing cloud collisions, differential shock heating, and cooling throughout. A wide range of relative gas phases and line-of-sight velocity distributions are found in the bridges, depending sensitively on initial disk orientations and the resulting variety of cloud collision histories. Most cloud collisions can occur promptly or persist for quite a long duration. Cold and hot phases can largely overlap throughout the bridge or can be separated into different parts of the bridge.

\end{abstract}



\section{Introduction: Splash Bridges}

Splash bridges are gas structures produced from the direct collision of two or more gas-rich disk galaxies, as seen in the well-known Taffy galaxies \citep{condon93}. In such collisions, up to half of the system's total gas can be pushed out into the bridge, which can persist for timescales of tens to hundreds of millions of years, as shown in previous works and other simulations \citep{10.1093/mnras/stu2713,10.1093/mnras/stw3011,vollmer12,vollmer2021,yeager19,Yeager_2020MNRAS.492.4892Y,Yeager_2020}. This large-scale disruption and rearrangement of the disk interstellar medium (ISM) mix the metals of the galaxy disks, lead to a rapid evolution of the gas phase balance, and can drastically change the star formation history of the remnants. Depending on the relative velocity of the collision, the two galaxies may merge promptly, merge with considerable delay, or not merge at all. In the case of a prompt merger, the bridge dynamics may be less important compared to other processes in the merger evolution. However, in cases of delayed merger, and especially in cases with no merger, the delayed accretion out of the bridge will profoundly affect disk evolution, and heating processes within the bridge may give birth to or add to a circumgalactic medium \citep{tumlinson2017circumgalactic}.

Though rare in the current universe, splash bridges provide a unique opportunity to study gas at a particular moment in a gas-rich galaxy collision when gas is drawn into a highly multi-phase medium driven temporarily by cloud-cloud collisions and turbulence. Because this material is spread out and removed from the obscuring parent disks, the bridge environment can be much cleaner and easier to study turbulent dissipation processes. The heating and cooling of the gas and the mixing of gas phases can be studied away from the confusing effects of starbursts or AGN-driven outflows. The central theme of this paper will be the evolution of cloud-cloud collisions in the large-scale streams of bridge gas pulled out of the two galaxies and how these collisions affect the balance of thermal phases, potentially teaching us general lessons about ISM dynamics in galaxy collisions, mergers, and accretion events.

Direct disk-disk encounters were evidently much more common in the early stages of large galaxy formation (see Fig. 6 of \citet{10.1093/mnras/staa1923}), as shown by the videos from the FIRE, EAGLE, and Illustris cosmological structure formation models \citep{2014MNRAS.445..581H,mcalpine2016eagle,vogelsberger2014introducing}. \footnote{https://fire.northwestern.edu/ \citep{fire}} \footnote{https://icc.dur.ac.uk/Eagle/ \citep{eagle}} \footnote{https://www.illustris-project.org/ \citep{illustris}}. To date, however, there have been no systematic studies in these models of these collisions, which are a subset of merger trees. Models and studies of nearby systems help us understand the effects of these collisions on galaxy evolution at high redshift. Since cosmological models generally have limited particle and mass resolution, they cannot resolve the phase evolution and accretion history as well as special-purpose models like those described below. Because of the lack of resolution, cosmological simulations are not well suited to studying splash bridge structures and cannot provide insight into the evolution of the unique environment produced in gas-rich disk-disk collisions. Zoom-in models can overcome this limitation but are complex and require large amounts of computer time, making it expensive to run many cases. Dedicated simulations are required to uncover the evolution of splash bridge structures in interactive gas-rich disk-disk collisions. Given these general considerations, we next summarize extant observations of splash systems.

The bridge between the Taffy galaxies (UGC 12914/5) was first detected in neutral hydrogen and radio continuum emission by \citet{condon93}. This prototypical splash bridge has now been observed extensively in many different wavebands, from the visible \citep{1987ApJ...320...49B,2019ApJ...878..161J} to the near, mid-, and far-IR \citep{1999AJ....118.2132J,peterson12,2018ApJ...855..141P}, and X-rays \citep{appleton15}. In addition to the $\sim 4 \times 10^9 M_{\odot}$ of HI originally discovered in the bridge \citep{condon93}, early CO observations of the Taffy system revealed large quantities of molecular gas, ranging from 2-10$ \times 10^9 M_{\odot}$ depending on the assumed ratio of X$_{CO}$, the ratio of the intensity of the CO line to the total molecular gas column \citep{braine03,gao03,2007AJ....134..118Z}. More recent observations of the bridge in CO emission made with both the PdBI and ALMA \citep{vollmer2021,Appleton_2022} have revealed that the gas lies in a tangled web of filaments with evidence of locally high-velocity dispersion, suggesting strongly turbulent gas. The high-resolution ALMA observations also showed that, except in one region of the bridge, the dense molecular gas is highly deficient in star formation.

Evidence of the bridge containing a multi-phase medium suffused with shocks and turbulence comes from many sources. \citet{peterson12} showed using {\it Spitzer} that, like the giant filament in the IGM of Stephan's Quintet, the bridge emits strongly in pure rotational lines of molecular hydrogen. The lack of star formation and associated PAH emission, combined with strong mid-IR H$_2$ lines, strongly suggested shocks as the heating source for the warm H$_2$. Follow-up observations with {\it Herschel} showed strong [CI] and \textnormal{[CII]} emissions from warm gas in the bridge that could not be explained by star formation heating. Further direct evidence for shocks in the bridge also came from optical IFU spectroscopy \citep{2019ApJ...878..161J}, which clearly showed faint emission line ratios consistent with models of 200-300 \kms shocks. Finally, soft X-ray emission was found in the bridge consistent with much faster shocks in a very hot bridge component \citep{appleton15}. These observations point to the bridge being heated by shocks moving at different speeds and through different components, ranging from a few \kms for the warm molecular hydrogen to 500-600 \kms for the hot X-ray gas.

Returning to the recent ALMA observations of the Taffy bridge, \citet{Appleton_2022} showed that the cold component of gas consisted of filaments of gas crisscrossing the region with locally large (70-100 \kms) velocity dispersion measured on the scale of 60-90 pc. Over a wide range of possible values for X$_{CO}$, the gas was found to be virtually unstable. Using H$\alpha$ observations taken at the same resolution as ALMA, it was shown that the high surface densities of gas were strongly deficient in star formation over much of the bridge compared with normal galaxies. It was suggested that cloud-cloud collisions turbulently heated the gas within the bridge. Indeed, such cloud-cloud collisions were predicted by models presented in a previous paper in the same series as this one by \citet{Yeager_2020MNRAS.492.4892Y} and by \citet{vollmer12,vollmer2021}. However, these models did not treat the multi-phase nature of the medium. In the era of the James Webb Space Telescope and its ability with MIRI to detect and map both the ionized gas and warm molecular gas at a comparable resolution to ALMA, we are truly entering a new phase in the observational understanding of the interaction of different gas phases in collisional systems. Part of the motivation of the current paper is to help to provide some additional insight into how the different gas phases evolve and interact in a turbulent environment.

Another splash bridge in UGC 813/816 is similar to the Taffy, containing a molecular bridge and connected by radio continuum emission \citep{2002AJ....123.1881C, Braine2004}. Additionally, at least partial \HI\ bridges have been detected in several colliding ring galaxies, including the Cartwheel \citep{hstcartwheel1994AAS...185.6806B,hstcartwheel1996AJ....112.1868S,higdon1998optical,combesmodel2001Ap&SS.276.1141H,2010MNRAS.403.1516S}, Arp 147 \citep{1992ApJ...399L..51G,fog11}, VII Zw 466, and the Sacred Mushroom \citep{1994sacredmushroomstruck,VIIZW4661996ApJ...468..532A}. With the exception of the Cartwheel, none of these systems have been observed in the same detail as the Taffy bridge. Given the fact that they result from the same type of collision, we expect splash bridges to be nearly as common as ring galaxies.

Evidence of cross-fueling between the northern galaxies and the southern galaxy of Arp 194 suggests the presence of a splash bridge with an estimated age of several hundred million years in that system \citep{arp194marziani2003}. Unlike most candidate splash bridges, the Arp 194 bridge contains a stream of star-forming knots tens of kiloparsecs in length. This may indicate that it is a hybrid tidal/splash bridge. The systems UGC 8335, Arp 240

, NGC 4485, Arp 148, and II Zwicky 96 may also contain gas bridges between the two disk galaxies. However, these systems have not been observed in a wide range of wavelengths, so the suggestion that they are splash bridges is based on optical morphology and observations in the HI Rogues Gallery.

Our previous work showed that splash bridges can persist for at least 100 Myr. Earlier work also modeled the collision of gas-rich galaxy disks and found a wide range of collision parameters that can lead not only to long-lived splash bridges but also to the formation of entirely new galaxies from the remnant, e.g., \citet{2011ApJS..194...47V}. The main reason that few splash bridges have been discovered and studied is that they generally lack a bright stellar component and are extremely faint for much of their lifetime in the optical, e.g., \citep{Yeager_2020}. The first splash bridges were discovered through \HI{} observations. Observations of the Taffy bridge have shown an extensive molecular component of the Taffy bridge, which suggests the presence of substantial molecular gas in other splash bridges. The molecular material may be stripped from the host galaxies and/or produced behind the strong shocks generated in the gas-rich disk-disk collisions, much as proposed by \citet{gui09} for the Stephan's Quintet system. We present new results relevant to this question below. These results support the proposition that cold molecular gas should be a common constituent in most splash bridges. This also highlights the question of why extensive star formation is not present in these structures, which will be addressed below.

Earlier papers in this series showed that the direct collision of two gas-rich disks produces a cascade of shocks that span a wide range in strength. This range extends up to a couple of orders of magnitude higher than those produced in the evolution of a typical swing bridge. This large range of shock strengths also leads to a large range of gas temperatures and densities, and to large offsets in the distribution of gas phases in splash bridges. This is seen in the Taffy galaxies and in many of the models described below. The kinematics of these bridges can also distinguish them from swing bridges and constrain the collision parameters. The evolution of the thermal effects and kinematics in the first several Myr after impact are the primary focus of this paper. In Section 3, we examine these thermal effects, and in Section 4, the kinematic effects, in a variety of direct 'splash' collisions. First, in the next section, we summarize the computational methods, highlighting differences from previous papers in this series.

\section{Methods}
\setcounter{subsection}{0}
\label{sec:methods}

\subsection{Summary}

The basic parameters of the galaxy disks in these models were inspired by the observed quantities of the Taffy Galaxies, UGC 12914 and UGC 12915. The northern disk in the Taffy, UGC 12915 is slightly smaller in apparent diameter than the southern disk of UGC 12914. For the sake of discussion in this paper, we use the abbreviations `G1' to refer to the southern disk and `G2' to refer to the northern disk.

The disk hydrodynamics were modeled using the sticky particle code described in \citet{yeager19}. The effects of various collision parameters on splash bridge morphology in these models were detailed in \citet{Yeager_2020MNRAS.492.4892Y}. An analysis of the star formation properties of splash bridges, using cloud collapse rules, was provided in \citet{Yeager_2020}.

To summarize some key aspects of the model code from our previous papers: the gravitational potential of each galaxy is modeled by a three-component Miyamoto-Nagai (MN) disk potential and a halo as described by \citet{hernquist1990}. The positions of the potential centers of each galaxy component move under the influence of all the others, as do the gas particles in each galaxy. The Matlab integration package ODE23 is utilized to solve equations of motion, with data output every 100,000 years for a total simulation time of 150 Myr. A nearest neighbor search is used to determine when gas particles are within 50 pc of each other and assumed to collide. Collisions between gas particles are treated as fully inelastic, triggering shocks in both particles. The downstream gas temperature, density, and shock velocity in each gas cloud are determined from the Rankine-Hugoniot jump conditions. Once half of the cloud crossing time has passed, shock heating and cooling are applied to the gas. time for each shock. One-half of the crossing time has passed shock heating and cooling are applied to the gas. Following is a compilation of the methods from our previous works in more detail.

The model initializations are summarized in Table \ref{table:modelparameters}. We do not provide figures for every one of these models in this paper. We will discuss various aspects of these models with a focus on the 0\degree{}, 45\degree{}, and 90\degree{} inclination models, which span the range of the model grid parameters. Thus, to some degree, the properties of the other models can be interpolated from these representative models.

\subsection{Initial disk configurations}
\begin{figure}
 \includegraphics[width=0.4\textwidth]{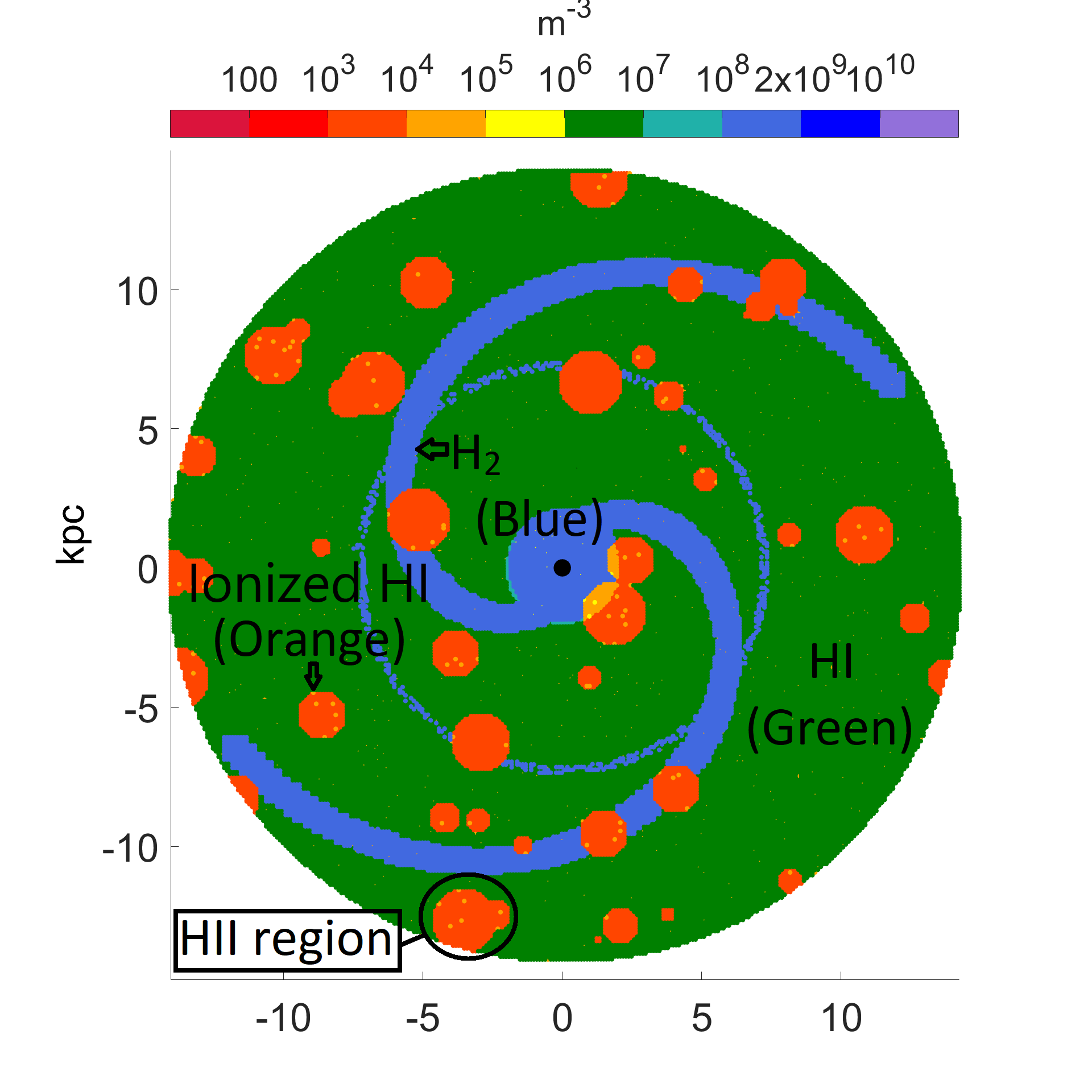}
 \caption{Example initial ISM distributions and densities. The circular regions in red are the HII hot gas 200 pc to 2 kiloparsecs in size; in blue, the spiral arms contain \hh{} material, and green makes up the most abundant component of neutral Hydrogen. Not easily seen are 100 pc gas clouds of light blue and orange scattered randomly through the disc, representing the cold neutral and warm neutral ISM phases. }%
 \label{fig:initialdisc}
\end{figure}

\indent An example of an initial disc is shown in \cref{fig:initialdisc}.  The initial separation between galaxy centres (z) is set to -15 kpc for every run. In all cases, the smaller G2 galaxy is placed below G1 in the vertical (z) direction.  An initial velocity for G2 is set to 400 $km/s$ in the z direction toward G1.  The disc plane of galaxy G2 is inclined relative to the x-y plane by different amounts in different runs.  The disc gravitational potential of G2 is also inclined to match the initial gas disc of G2.  

\indent Each run starts approximately 25 Myr before the galactic centres reach closest approach.  This allows for gravitational effects to begin warping the discs prior to the collision.  Initially, the two discs are given: a separation in x,y,z, a velocity in the z direction, and an inclination measured from the x-y plane about the x-axis (same axis of offset), unless otherwise specified.  The offset is defined as how far apart the galactic centres are initially placed.  

The gas in each galaxy is initialized into five phases. The density of the gas clouds is scaled following a $1/R$ relationship as well as a scaling depending on the initial ISM phase. The density of the \HI{} is determined by \Cref{eq:densityofdisk}

\begin{equation}
  \rho (R) = \frac{8\times10^2}{R + a} \text{ cm}^{-3}
  \label{eq:densityofdisk}
\end{equation}

where \( R \) is the radius from the center of the gas disk and \( a \) is the disk scale length, in kiloparsecs. The scale lengths are 6.3 and 3 kiloparsecs for G1 and G2. The peak value of \SI{8e2} \dunit{} is informed by the peak density calculated from the density formula derived from the Miyamoto-Nagai gravitational potential used for each disk, as found in \citet{smith15}. The resulting \HI{} cloud densities are in a range expected for \HI{} gas clouds found in the Taffy Galaxy UGC 12914. The central density of \HI{} gas in G1 is 13.3 \dunit{} and in G2 is 9.26 \dunit{}, with respective 'edge' densities of 2.32 \dunit{} and 1.28 \dunit{}. The cloud densities for the other four ISM phases are simply scaled from the \HI{} gas density. The scaling factors for \hh{}, CNM, and the two ionized HI phases are 100, 10, 0.02, and 0.002, respectively. \Cref{fig:initialdisc} is an example distribution of these gas phases within a simulated galactic disk. The five gas phases are given initial temperatures of 50, 100, 5000, \SI{5e5}, \SI{5e6} Kelvin, respectively. The gas phases do not evolve through heating or cooling prior to their first collision. The ISM is initialized to represent a typical Sc galaxy, so it is assumed there are internal processes in each phase that maintain the relative temperatures and densities over the approximate 10 Myr before they experience a collision.

\begin{table}
\begin{adjustwidth}{-0.5cm}{}
\centering
\begin{tabular}{|c|c|c|c|c|c|c|c|c|c|}
\hline
X & Y & Z & $V_X$ & $V_Y$ & $V_Z$ & $V_{\text{impact}}$ & X tilt & Y tilt \\
\hline
0 & .5 & -15 & 0 & 0 & 400 & 938 & 0 & 0 \\
0 & .5 & -15 & 0 & 0 & 400 & 884 & 0 & 20 \\
0 & .5 & -15 & 0 & 0 & 400 & 919 & 0 & 45 \\
0 & .5 & -15 & 0 & 0 & 400 & 869 & 0 & 65 \\
0 & .5 & -15 & 0 & 0 & 400 & 934 & 0 & 90 \\
0 & 10 & -15 & 0 & 0 & 400 & 642 & 0 & 0 \\
0 & 10 & -15 & 0 & 0 & 400 & 621 & 0 & 20 \\
0 & 10 & -15 & 0 & 0 & 400 & 638 & 0 & 45 \\
0 & 10 & -15 & 0 & 0 & 400 & 625 & 0 & 65 \\
0 & 10 & -15 & 0 & 0 & 400 & 649 & 0 & 90 \\
0 & 10 & -15 & 0 & 0 & 400 & 654 & 90 & 0 \\
0 & -10.6 & -10.6 & 0 & 282 & 282 & 900 & 0 & 0 \\
0 & 20 & 20 & 0 & -350 & -350 & 1018 & 0 & 0 \\
\hline
\end{tabular}
\caption{Initial Parameters for all of our Simulated Galaxy Disk-Disk Collisions. The first three columns, X, Y, and Z in kiloparsec, are the position of G2 relative to G1. $V_{X}$, $V_{Y}$, and $V_{Z}$ are the initial velocities in \kms{} of G2. $V_{\text{impact}}$ in \kms{} is the magnitude of the net velocity at which the gas disks collide. This does not include the counter-rotation velocities that also contribute to the gas cloud--cloud collision velocities. G1 is initially at rest in all runs. X tilt and Y tilt are angles in degrees at which the gas disk of G2 is inclined relative to the X- and Y-axis, respectively.\label{table:modelparameters}}
\end{adjustwidth}
\end{table}

\subsection{Gas Phases}
\label{sec:gasphases}
The interstellar gas in galaxy disks has a wide range of temperatures and densities. In setting up the initial conditions in our models, we do not attempt to represent this continuum of thermal phases. We certainly do not attempt to establish a statistically steady state among them since this would require including several uncertain feedback processes. Instead, we choose five baseline thermal phases, which provide a simple approximate view of the full spectrum of possible states. Since disk-disk impact begins shortly after the start of the models, we also do not try to evolve these phases before impact. The gas phases are primarily defined by their characteristic temperatures, and each cloud particle is initially assigned a phase and the corresponding temperature. Initially, each phase has a characteristic density assigned at a scale radius, so it is in pressure balance with the other phases locally. The density of each particle is scaled with radius according to the overall gas density profile. These initial phases are selected to roughly correlate with the emissions in important observational wavebands, though they do not directly correspond to observables. Because these model states are artificial, we give them names related to corresponding real phases. After the disk-disk impact, we track the continuous thermal transitions between each particle due to shock heating and various cooling processes in the post-shock gas. 

\begin{figure}
  \begin{center}
   \includegraphics[width=.5\textwidth]{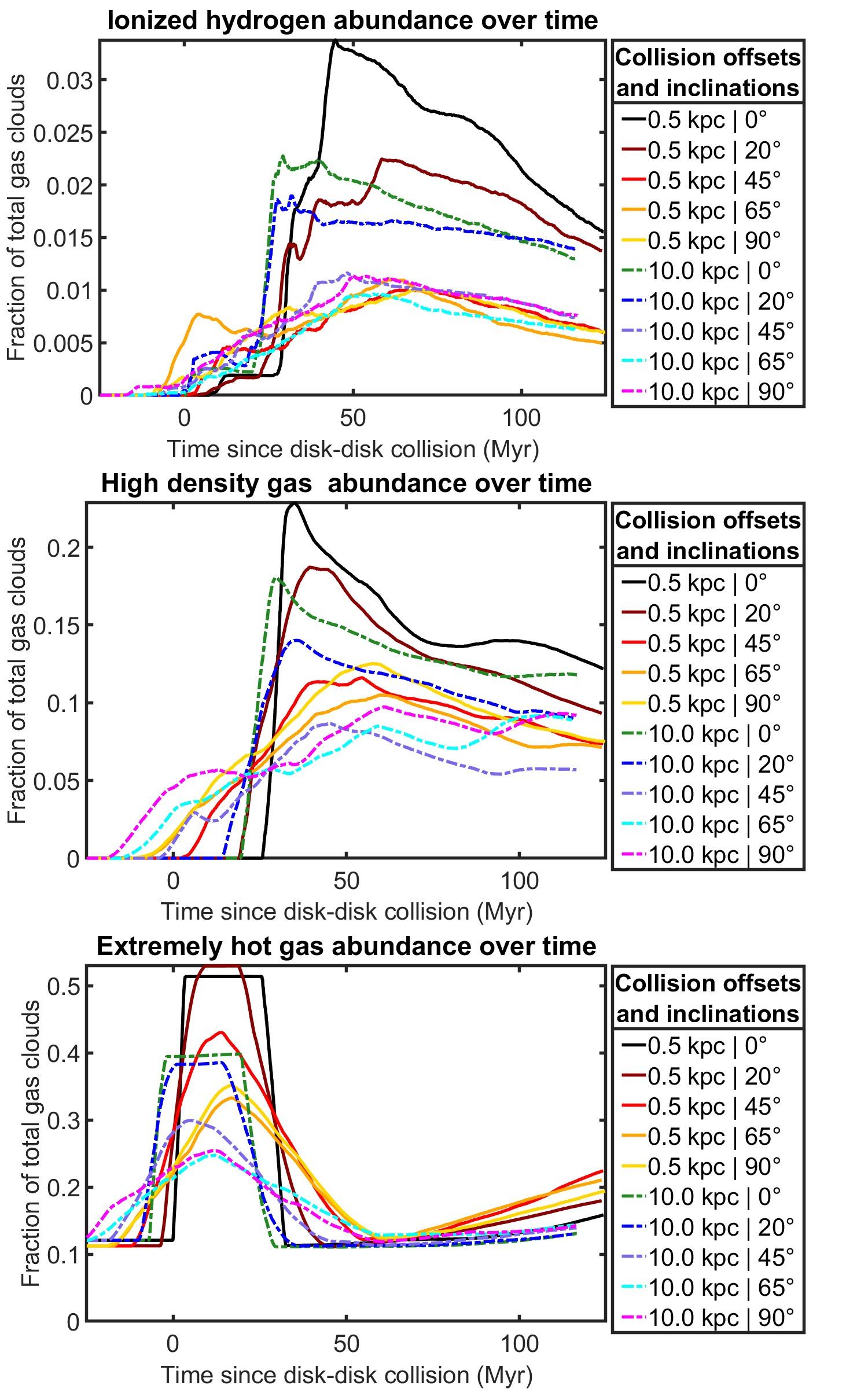}
\caption{The time evolved-gas fraction for three phases of ISM defined in Section \ref{sec:methods}. The color indicates the type of disk-disk collision as shown in the legend, solid lines are all 500 pc offsets, and dashed lines are ten-kiloparsec offsets. \textit{Initially dense gas} is not shown as it is nearly a constant $10.6\%$ fraction with time for all collisions.}  %
\label{fig:abundances}
 \end{center}
\end{figure}

Five phases of ISM are defined here, chosen for their particularly interesting cooling processes - or lack thereof. Not to be confused with the five temperature and density ranges used to initialize the galaxy disks discussed in section \ref{sec:gasphases}. Though rare, as told by the results later in the paper, a single gas cloud can possibly fall into multiple of these categories, e.g. the cloud ends up at high density by also high temperatures due to repeated shocks. Listed are the five phases, along with a sub-list of the algorithmic cooling active emission processes.
\begin{enumerate}
  \item \textit{Extremely hot} gas; gas with a temperature greater than one million Kelvin.
  \begin{enumerate}
    \item Bremsstrahlung
    \item dust blackbody
    \end{enumerate}
  \item \textit{Ionized hydrogen}; gas with a temperature between six thousand and ten thousand Kelvin. [CII] can occur in this gaseous phase, but [CII] emission is only relevant when the temperatures and densities reach a level below the critical density.
  \begin{enumerate}
    \item hydrogen line
    \item helium line
    \item molecular hydrogen line
    \item silicon line
    \item oxygen line
    \item dust blackbody
    \item \textnormal{[CII]} fine structure line*
  \end{enumerate}

  \item \textit{High density gas}; gas with a density greater than \SI{2e3} \dunit{}. This is the density at which \textnormal{[CII]} fine structure cooling is suppressed.
  \begin{enumerate}
    \item molecular hydrogen line
    \item silicon line
    \item oxygen line
    \item dust blackbody
  \end{enumerate}

  \item \textit{Initially dense gas}; The initially dense gas is designated to track the evolution of the densest gas - the (\hh) phase, before the disk-disk collision. Since the transfer of momentum between gas clouds is the dominant physical process that produces splash bridges, the initial density being an order of magnitude greater than all other gas phases will produce distinctly different kinematics post-collision. Unlike the other four phases, no further restrictions are placed on the gas cloud's evolved temperature or density.

  \item \textit{Cool gas}; temperature above 150 K and less than 5,000 K. This phase of gas makes up the majority of the gas in the disks before the collision and is the end state of gas that does not under go repeated cloud-cloud collisions. This phase is not specifically isolated in the plots below as the this phase accounts for all gas not existing within one of the previously mentioned four phases. The cold gas phase also closely matches closely with the overall gas profiles which are represented by the black lines in \Cref{fig:0inc-rotx0y0z0} to \Cref{fig:90inc-rotx0y0z90}.
 \begin{enumerate}
    \item hydrogen line
    \item molecular hydrogen line
    \item \textnormal{[CII]} fine structure line
    \item oxygen line
    \item silicon line
    \item dust blackbody
 \end{enumerate}

\end{enumerate}
The abundances of these ISM phases over time are shown in \Cref{fig:abundances} for the ten models outlined in Table \ref{table:modelparameters} and discussed below.

For the present paper, we have updated our cooling algorithm from that used in \citet{Yeager_2020MNRAS.492.4892Y} to include approximate cooling functions for oxygen, silicon, and molecular hydrogen line cooling. Again, the gas in these models is initialized into five distinct phases, each determined by its initial temperature and density. The initial distribution of these phases in the disks is discussed in section \ref{sec:gasphases} and detail in \citet{yeager19}. The gas of each phase is initially at pressure equilibrium, with the other phases determined by the gas temperature (T) and volume density (n). The five phases represent the interstellar medium (ISM) found in a typical gas-rich Sc-disk galaxy. The most abundant phase fills about 90\% of our gas disks. It is modeled as a warm hydrogen gas with temperature 5,000 K and $n_{\HI{}}$ densities on the order of 10 \dunit{} in the central disk, and that density falls off exponentially outwards. 

The two coldest phases are meant to represent dense clouds and more diffuse molecular material and are given initial temperatures of 50 K and 100 K and volume densities of order $100n_{\HI{}}$ and $10n_{\HI{}}$, respectively. This gives the \hh{} gas clouds densities on the order of $10^{2}-10^{3}$ \dunit. These \hh{} phases account for approximately 10\% of the total gas by simulation cell count within our galaxy disks. The remaining volume of our gas disks is populated with ``hot diffuse" gas. Initially, this gas consists of two hot phases, which are randomly assigned either a temperature of T$=$\SI{5e5} K and corresponding density scaling factors of $10^{-2}n_{\HI{}}$ or T$=$\SI{5e6} K and density scaling of $10^{-3}n_{\HI{}}$.

\subsection{Cloud-Cloud Collisions}

Initially, a Cartesian x-y grid is projected onto both galactic discs, crossing the simulation volume. Cells with matching (x,y) coordinates in the two disc projections directly collide and stick together completely inelastically at their contact face. Cells at the edges of each disc are modeled with one curved face instead of being perfectly rectangular. For the models discussed in this paper, the grid cell length is set to 100 pc.

When a collision occurs, a contact discontinuity forms. On each side of this discontinuity, the gas remains in distinct phases without mixing. Shock waves are generated at the contact discontinuity and propagate through each cloud. Post-collision, the gas elements on either side of the discontinuity move together at their center of mass velocity, which is determined by conserving momentum across the discontinuity. This center of mass velocity serves as the initial condition for integrating the motion equations of the merged gas elements, referred to as particles or clouds. We use the Matlab ode23 integration package to solve these second-order Newtonian motion equations by decoupling them.

\begin{equation} 
\dv{q}{t}= v_{q}
\end{equation} 

\begin{equation} 
\dv{q}= - \dv{q} \phi(x,y,z),
\end{equation}

Here, $q$ represents a coordinate direction, $v_q$ is the corresponding velocity, and $\phi$ is the total gravitational potential.

Gas particles in both galaxies are influenced by all potentials and are evolved with a single timestep across the grid to minimize integration errors.

\subsection{Shocks and Heating}

Shocks are generated at the contact discontinuity where colliding clouds meet, propagating through each cloud.

In a steady-state, inviscid, non-accelerating, ideal gas scenario, we use the following equations:

\begin{equation} 
\epsilon = \frac{3 k_{B} T}{2 m},
\end{equation} 

\begin{equation} 
P = \frac{2}{3} \rho \epsilon = n k_{B} T,
\end{equation} 

where $\rho$ is the local gas density, $n$ is the number density, $P$ is the pressure, $\epsilon$ is the specific energy, $k_B$ is the Boltzmann constant, and $m$ is the mean particle mass.

The Rankine-Hugoniot jump conditions, assuming no heating or cooling ($\Gamma = \Lambda = 0$) across shocks, are:

\begin{equation} 
\rho_{up} v_{up} = \rho_{d} v_{d},
\end{equation} 

\begin{equation} 
\rho_{up} v_{up}^2 + P_{up}= \rho_{d} v_{d}^2 + P_{d},
\end{equation} 

\begin{equation} 
\frac{1}{2} \rho_{up} v_{up}^2 + \epsilon_{up} + \frac{P_{up}}{\rho_{up}} = \frac{1}{2} \rho_{d} v_{d}^2 + \epsilon_{d} + \frac{P_{d}}{\rho_{d}},
\end{equation}

with $v$ representing flow velocities. The subscripts up and d denote upstream and downstream gas flows in front of and behind the shocks, respectively (e.g., \cite{drainephysics}).

In areas without strong shocks, heating is minimal. We focus on strong shocks as we are interested in gas that requires significant time to cool. For strong shocks, the following approximations hold:

\begin{equation} 
c = \sqrt{\gamma \frac{k_{B} T}{\mu m_{h}}},
\label{eq:soundc}
\end{equation}

\begin{equation} 
a = \frac{v_{up}}{c_{up}} = \sqrt{\frac{\rho_{up} v_{up}^2}{\gamma P_{up}}} \ (Mach \ number),
\end{equation}

\begin{equation} 
\frac{\rho_{d}}{\rho_{up}} = \frac{v_{up}}{v_{d}},
\end{equation}

\begin{equation} 
P_{d} = \frac{3 \rho_{up} v_{up}^2}{4},
\end{equation}

\begin{equation} 
T_{d} = \frac{3 m v_{up}^2}{16 k_{B}},
\end{equation}

where $a$ is the Mach number, $c$ is the speed of sound, and $\gamma$ is the ratio of specific heats, assumed to be 5/3 for all shocks.

The jump conditions are applied in the shock's moving frame. The velocity of the gas behind the shock is assumed to be at rest relative to the cloud system's center of mass (see Figure 1 in \cite{yeager19}). This allows us to calculate the velocity of each shock relative to the center of mass. In extreme cases, post-shock temperatures can reach up to \SI{e7} K.

Cloud collisions are detected using a nearest neighbor search after each time step. This frequency of detection differs from previous code versions but is necessary due to the large disc-disc inclinations that prevent using a single grid to define colliding gas clouds. Collisions between gas clouds are treated as inelastic momentum-conserving collisions, similar to previous face-on models. Figure 1 in \cite{yeager19} illustrates the setup of shocks in a cloud-cloud collision, showing the contact discontinuity where shocks are generated and propagate back through their parent clouds.

Multiple gas clouds can find themselves within each other's radii over a time step. However, for computational efficiency, only the nearest gas clouds are considered colliding. For instance, if cloud A finds clouds B, C, and D within its collision radius at a time step, the code identifies C as the closest to A, recording a collision between A and C. Similarly, if B finds A as its closest neighbor, a collision between B and A is recorded in the same time step. For cooling computations, only the strongest shock is considered each time step.

Each cloud has an initial 'collision radius' equal to half the set resolution of the simulation, typically 100 pc unless specified otherwise. A 'cooling radius' updates as the cloud cools or expands adiabatically, separating cooling processes from kinematics and maintaining similar mass overlap in collisions. Collisions occur if one cloud's center is within another's radius, with only clouds having relative velocities above 40 \kms{} considered. This velocity limit prevents continuous collision registration and ensures disc rotation does not prematurely trigger collisions, preserving the coldest ISM phases. Clouds colliding below 40 \kms{} will only experience minor heating, insufficient to significantly impact the medium in this run.

\subsection{Cooling and thermal physics}

In the following paragraphs, we summarize and update several cooling formulations used in \citet{Yeager_2020MNRAS.492.4892Y} and give more details on the newly added processes. Bremsstrahlung, hydrogen, and helium line cooling are approximated through an algebraically well-fit function described in \citet{wang15}. \textnormal{[CII]} fine structure cooling, H$_{2}$, silicon, oxygen line cooling, and UV photo-heating are modeled by constant cooling rates. The magnitude of each cooling function and the density at which each form of cooling is suppressed are approximated as in \citet{drainephysics}. 

Shock parameters are calculated for every collision. Still, jump conditions are only applied if they produce a post-shock temperature greater than their cloud's temperature at half the crossing time. If a shock is strong enough, the cloud temperature is increased to the post-shock temperature, and the density and radius are updated corresponding to the shock jump conditions when the half-crossing time is completed. After a shock has completed half a crossing time, cooling and heating processes are allowed to begin within the cloud.

\Cref{eq:coolingfunction} below is from \citet{wang15} and is used to approximate the cooling function, with a maximum error of $5\%$, for optically thin cooling by hydrogen and helium emission, as well as Bremsstrahlung.

\begin{equation} 
\begin{aligned}
\Lambda (T) = \frac{{x_1} T^{x_2} + ({x_3} T)^{x_4} ({x_5} T^{x_6} + x_7 T^{x_7})}{1 + ({x_3} T)^{x_4}} + {x_9} T^{x_{10}}
\label{eq:coolingfunction}
\end{aligned}
\end{equation}\Cref{eq:coolingfunction} is fit with the constants from \Cref{tab:cfconsts} and is converted into a rate of temperature change by the relation $\frac{dT}{dt} = \frac{2 \Lambda(T) n^2}{3 k_{B} n_{i} }$, where $\Lambda(T)$ is the cooling function, T is temperature, $k_{b}$ is Boltzmann's constant, $n_{i}$ is the initial gas cloud number density and $n_{i}$ is the current gas number density. \Cref{eq:coolingfunction} is valid over the temperature range \SI{2e4} to \SI{e10} K and at densities of less than \SI{e12} cm\textsuperscript{-3}, which is well above the values required for our models.

\begin{table}
\begin{tabular}{| c | c | c | c |}
\hline
Constant & Value & Constant & Value \\
\hline
$x_1$ & $4.86567\times10^{-13}$ & $x_2$ & -2.21974 \\
$x_3$ & $1.35332\times10^{-5}$ & $x_4$ & 9.64775 \\
$x_5$ & $1.11401\times10^{-9}$ & $x_6$ & -2.66528 \\
$x_7$ & $6.91908\times10^{-21}$ & $x_8$ & -0.571255 \\
$x_9$ & $2.45596\times10^{-27}$ & $x_{10}$ & 0.49521 \\
\hline
\end{tabular}
\caption{Cooling function constants for \Cref{eq:coolingfunction}.}
\label{tab:cfconsts}
\end{table}

Energy is transferred from hot gas particles to dust grains at a rate governed by the collision rate between particles and dust grains. Initially, 1$\%$ of the mass of each gas cloud is assumed to consist of dust grains. Mean dust grain radii are computed in each cloud and can change through sputtering, which depends on the local gas temperature.

The mean radius of the dust grains ($a$) are evolved via sputtering as per \Cref{eq:sputteringradius} from \citet{drainephysics},

\begin{equation}
  \label{eq:sputteringradius}
  \dv{a}{t} = - \frac{10^{-6}}{(1 + T_{6}^{-3})} \frac{n_{h}}{(cm^{-3})}\ \frac{\si{\micro \meter}}{yr}.
\end{equation}

If grains sputter below a radius of 1.0 nm, dust grain cooling is halted. When a dust grain collides with gas atoms most of the kinetic energy is transferred to the dust grain. Dust grains cool extremely fast compared to galactic time scales, so it is assumed that the gas instantly radiates $0.8kT$ Joules of heat after each dust grain collision. The dust collision rate ($\dv{N_{col}}{t}$) is given by, 

\begin{equation}
  \label{eq:colrate}
  \dv{N_{col}}{t} = n_{h} n_{dust} \pi a^2 v_{prob},
\end{equation}

\begin{equation}
  \label{eq:dustrate}
  \dv{T}{t} = -0.8 \dv{N_{col}}{t} T.
\end{equation}

\noindent In \Cref{eq:colrate} $v_{prob}$ is the most probable velocity of a single ISM particle, $n_h$ is the number density of the ISM, $n_{dust}$ is the number density of dust grains, $a$ is the current grain radius and $T$ is the temperature of the local ISM. 

Clouds are allowed to expand adiabatically at a rate governed by the local speed of sound, which is dependent on gas temperature. Adiabatic expansion is halted if the pressure within the cloud is not at least greater than an assumed intergalactic medium pressure of $10^{2}$ cm$^{-3}$ K. This basement pressure is based on a pressure value (nT) of $10^{-4}$ cm$^{-3}$ and $10^6$ K, typical of the observed intergalactic medium, \citet{Nicastro2018}.

Adiabatic cooling occurs when clouds expand (assumed spherically) at the speed of sound (c) \Cref{eq:soundc} is the same speed at which the shocks propagated. A value of $\gamma = \frac{5}{3}$ is assumed and the molecular weight $\mu$ is calculated for a mixture of hydrogen and He gas at a 3 to 1 ratio, $m_h$ is the mass of hydrogen.

\begin{equation} \dv{T_{adb}}{t} = \frac{-3(\gamma-1) c T}{R}.
\label{eq:workbyadb}
\end{equation}

\noindent An intergalactic pressure of \SI{e2} \dunit K is assumed to surround every cloud, \citet{Nicastro2018}. When the internal pressure of an expanding cloud falls below this pressure threshold, adiabatic expansion is halted.

\textnormal{[CII]} fine structure, molecular hydrogen, oxygen, and silicon line cooling are all approximated by a constant cooling function $\Lambda_{const}$ valid for the restricted temperature range of $150 K < T < 20,000 K$ and at densities less than the corresponding critical densities. The values of the cooling constants are: 1) \SI{1e-26} for \textnormal{[CII]}, 2) \SI{1.1e-26} for oxygen, when n is less than \SI{4.9e4} \dunit$\ $ or \SI{1.1e-27} if n is greater than \SI{4.9e4} \dunit$\ $ but less than \SI{2.5e5}, 3) \SI{1e-28}{} for silicon \citep{drainephysics}, and finally, 4) \SI{1e-25} for molecular hydrogen \citep{Roussel_2007}. All of these are in units $erg*cm^3/s$.

The constant cooling functions are turned into a rate of temperature change via,

\begin{equation} \dv{T}{t} = \frac{-2 \Lambda_{const} n^{2} }{3*k_{b}*n_{i}}
\label{eq:constantcoolingfunction}
\end{equation}

\noindent where $\Lambda_{const}$ is the relevant cooling constant, $n_{i}$ is the initial gas cloud number density, n the current gas cloud number density, and $k_{b}$ is the Boltzmann constant.

A constant rate of UV heating is also included. This rate is approximately $\frac{1}{3}$ of the \textnormal{[CII]} cooling rate, $\Lambda_{UV} = \SI{9.461e-33} J m^{3} s^{-1}$ as per \citet{drainephysics}.

The total differential equation for the temperature is solved using the Matlab ode23 solver, and temperatures and densities of all gas clouds are saved at the same intervals as the kinematics.

\section{Results 1: Thermal phases, heating, and cooling}
\subsection{Bridge mass evolution}
\begin{figure}
  \begin{center}
   \subfigure[]{%
   \includegraphics[width=.5\textwidth]{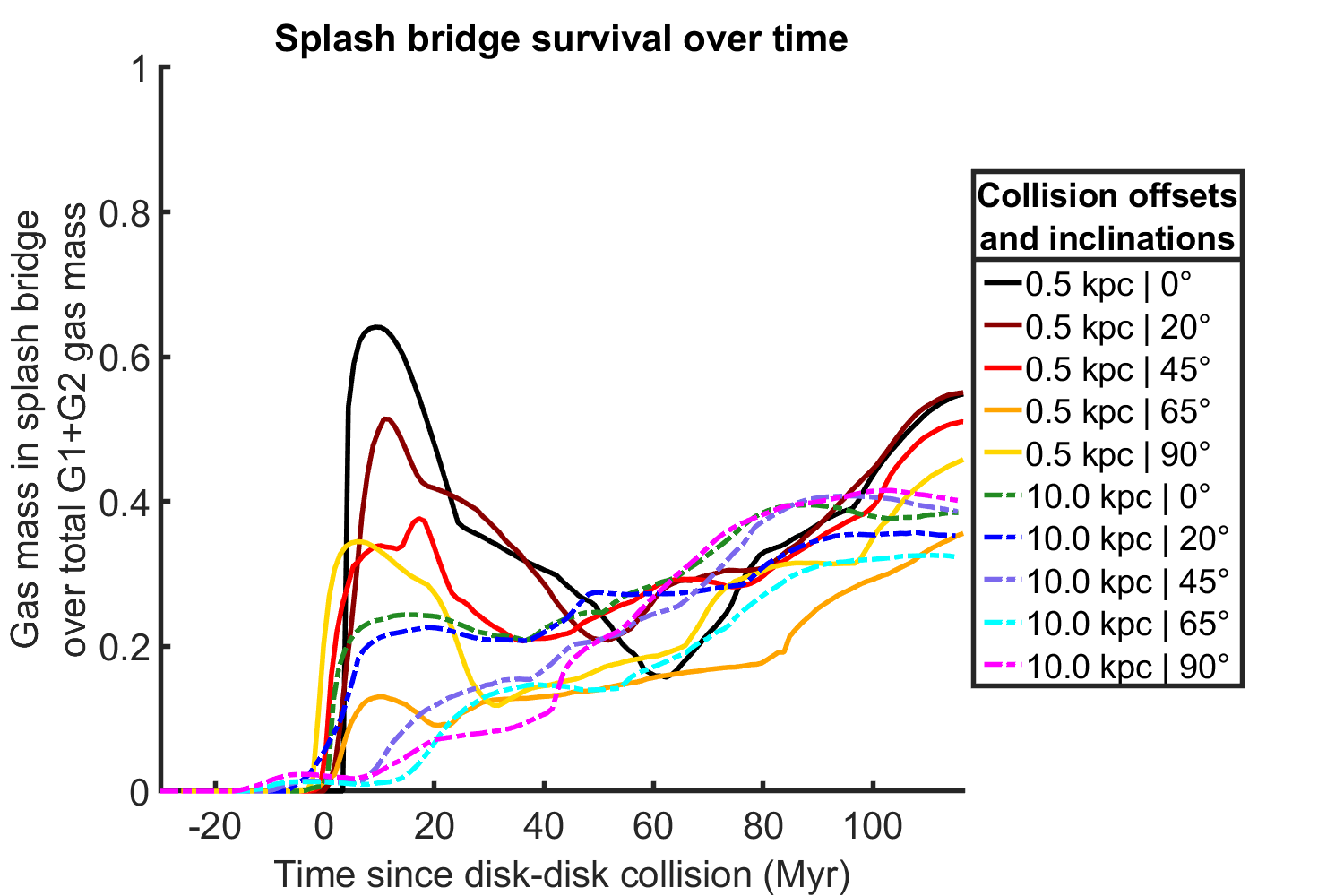}
   }
   \subfigure[]{%
   \includegraphics[width=.5\textwidth]{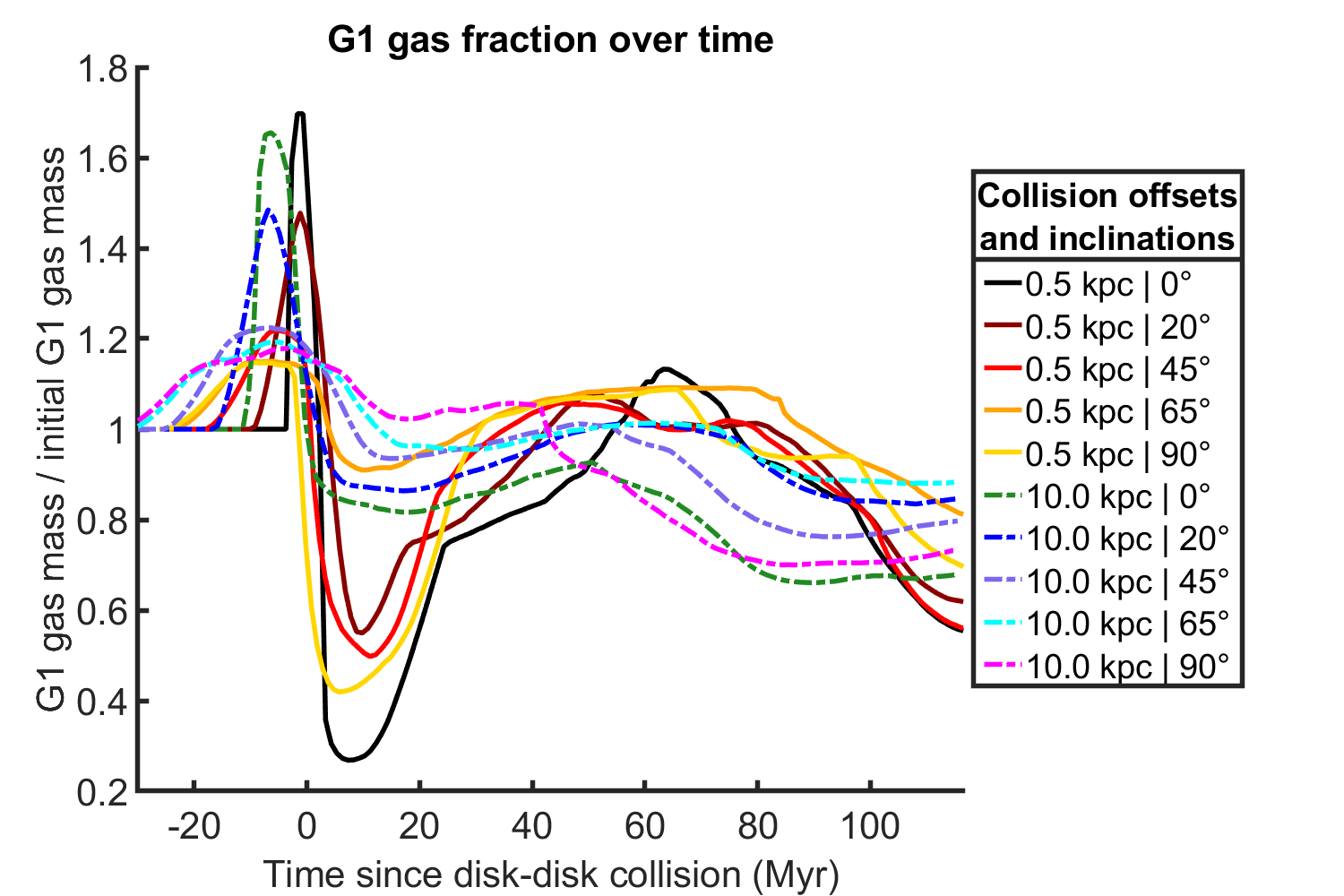}
   }
   \subfigure[]{%
   \includegraphics[width=.5\textwidth]{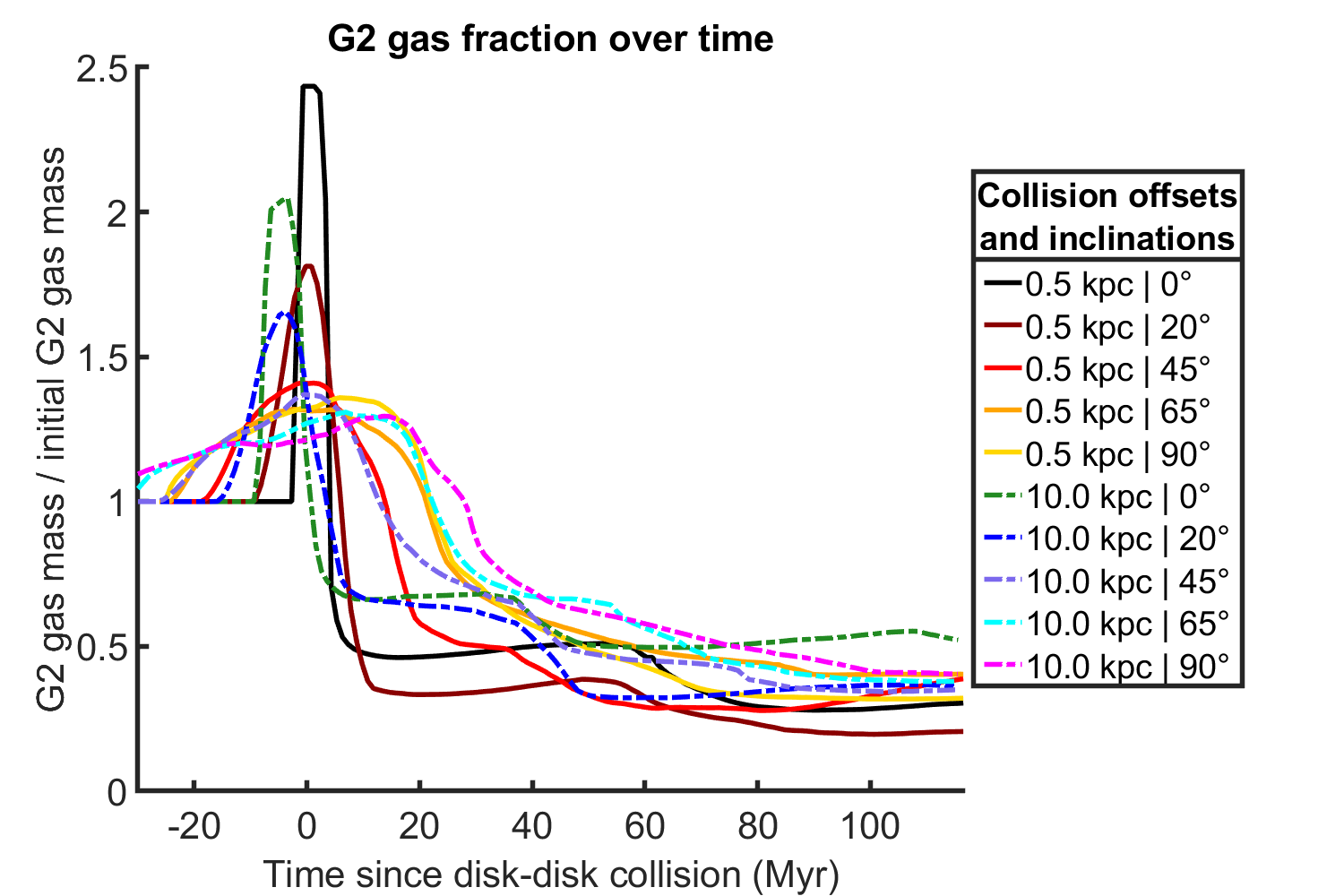}
   }
\caption{Panel (a) particles located at least two kiloparsec away from the plane of either galaxy disk and are within a radial distance of 15 kiloparsec of the disk center. Runs with a collisional offset of 500 pc are solid lines, and ten kiloparsec offsets are dashed lines. The color indicates the relative inclination between the colliding disks. Panel (b) G1 disk gas fraction over time, weighted to the initial gas mass in G1. Panel (c) G2 disk gas fraction over time, weighted to the initial gas mass in G2.} %
\label{fig:depletiontimes}
 \end{center}
\end{figure}

In \Cref{fig:depletiontimes}, we compare the amount of gas removed from the disks of each galaxy in all of the current models and find that most of the gas is removed in zero inclination and low offset collisions. Gas particles are determined to be within the splash bridge if they remain at least two kiloparsecs away from the gravitational plane of either galaxy disk and are within a radial distance of 15 kiloparsecs from the disk center. Runs with a central offset in the collision of 500 pc are shown as solid lines and ten-kiloparsec offsets are dashed lines. The color indicates the relative inclination between the colliding disks. As the relative inclination of the disks is increased, the fraction of gas removed drops from a peak of nearly 70$\%$ for the zero inclination, 500 pc offset disk-disk collision. Increasing the inclination between the gas disks both delays and reduces the maximum amount of gas that is splashed out of G1 and G2. For collisions with 10 kiloparsec offsets, the splash bridge gas fraction grows at a lower rate. For the 10 kiloparsec offset 0\degree{} inclination collision, the splash bridge still reaches a peak gas fraction of 57$\%$. For the zero inclination collisions, 35$\%$ of the galaxy system's gas is accreted back onto G1 and G2 over a 20 Myr period, between 20 to 40 Myr after the initial disk-disk collision. As seen in panels (b) and (c) of \Cref{fig:depletiontimes}, the accretion of splash bridge gas is mostly back onto G1, while G2 remains stripped of its initial gas. 

The total galaxy masses of each galaxy are \SI{4.4e11}{M\textsubscript{\(\odot\)}} and \SI{2.4e11}{M\textsubscript{\(\odot\)}} for UGC 12914/15 respectively (i.e., like the Taffy system determinations from \citealt{condon93}). The fact that G2 is half the gravitational mass of G1 results in a skewing of splash bridge gas accretion. We use this Taffy Galaxy mass ratio for all of our collision models. One can assume a mass ratio of unity would result in half of the splashed gas remaining bound to G1 and the other half bound to G2, resulting in equal accretion of the splash bridge mass onto both. Increasing the mass ratio of the galaxies will skew the accretion towards the more massive galaxy. A further complication is that the gas cloud densities within each galaxy may not be equal. This is the case for our colliding systems, with the densities of gas clouds in G1 being roughly 20$\%$ denser than those of G2. This helps explain the strong asymmetry of gas accretion onto the disk of G1 versus G2. Not only is G2 approximately half the gravitational mass of G1, allowing G1 to dominate in attracting splashed gas clouds, but the gas clouds of G1 also experience a more minor change in vertical momentum than the gas clouds of G2 due to the disparity in initial cloud densities.

On a timescale of 100 Myr, the fraction of gas lost from G1 remains between 10-20$\%$ for all collision inclinations. Low collisional offsets have more splash gas returned to the disk of G1 than high inclination offsets. The low offset - low inclination disk-disk collision results in the most gas splashed from both galaxy disks, and in our simulation, G1 is able to accrete the lion's share of the splashed gas. For the first 10 Myr following the disk-disk collision, G1 loses up to 40$\%$ of its gas content. But both gas splashed from G1 and G2 is accreted back onto G1 within 20 Myr following the disk-disk collision. The slightly smaller disk galaxy G2 retains less than 50$\%$ of its original gas mass. 

Inclination plays a role in the rate at which gas is removed from G2 but does not seem to strongly affect the total gas removed 100 Myr after the disk collision. In low inclination cases over 50$\%$ of the G2 gas, mass is removed within 10 Myr, but when G2 is inclined 90\degree{} relative to the disk of G1, it takes over 60 Myr to lose 50$\%$, and 80 Myr before the initial disk-disk collision is entirely over. For the G2 disk, large collision offsets result in more gas retention (50$\%$ original amount) at 100 Myr, and low offsets lead to about 40$\%$ of G2's gas being retained. This is the opposite of what is seen in G1.

In none of our runs do the disk galaxies accrete all of the splash bridge gas by the end of the 150 Myr simulation, suggesting splash bridge structures can be relatively long-lived and increasing the probability of observing them. We find that for the collision parameters we have covered, the 500 pc offset, 0\degree{} inclination model has the highest fraction of gas re-accreted onto both galaxies, with a bridge gas fraction of 10\% at 150 Myr. It also has the highest gas fraction `splashed' from both galaxies peaking at 70\%. Increasing the relative disk inclination shortens the delay in re-accretion down to no delay for disk-disk collisions with $90\degree{}$ inclinations.

\subsection{Dominant Cooling Processes}

\begin{figure}
  \begin{center}
    \subfigure[]{%
      \includegraphics[width=0.2\textwidth]{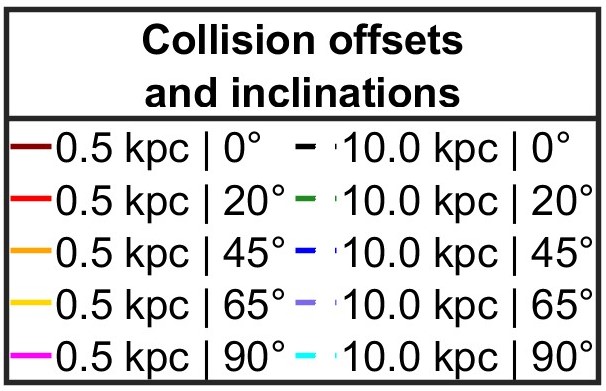}
    } \\
    \subfigure[]{%
      \includegraphics[width=0.39\textwidth]{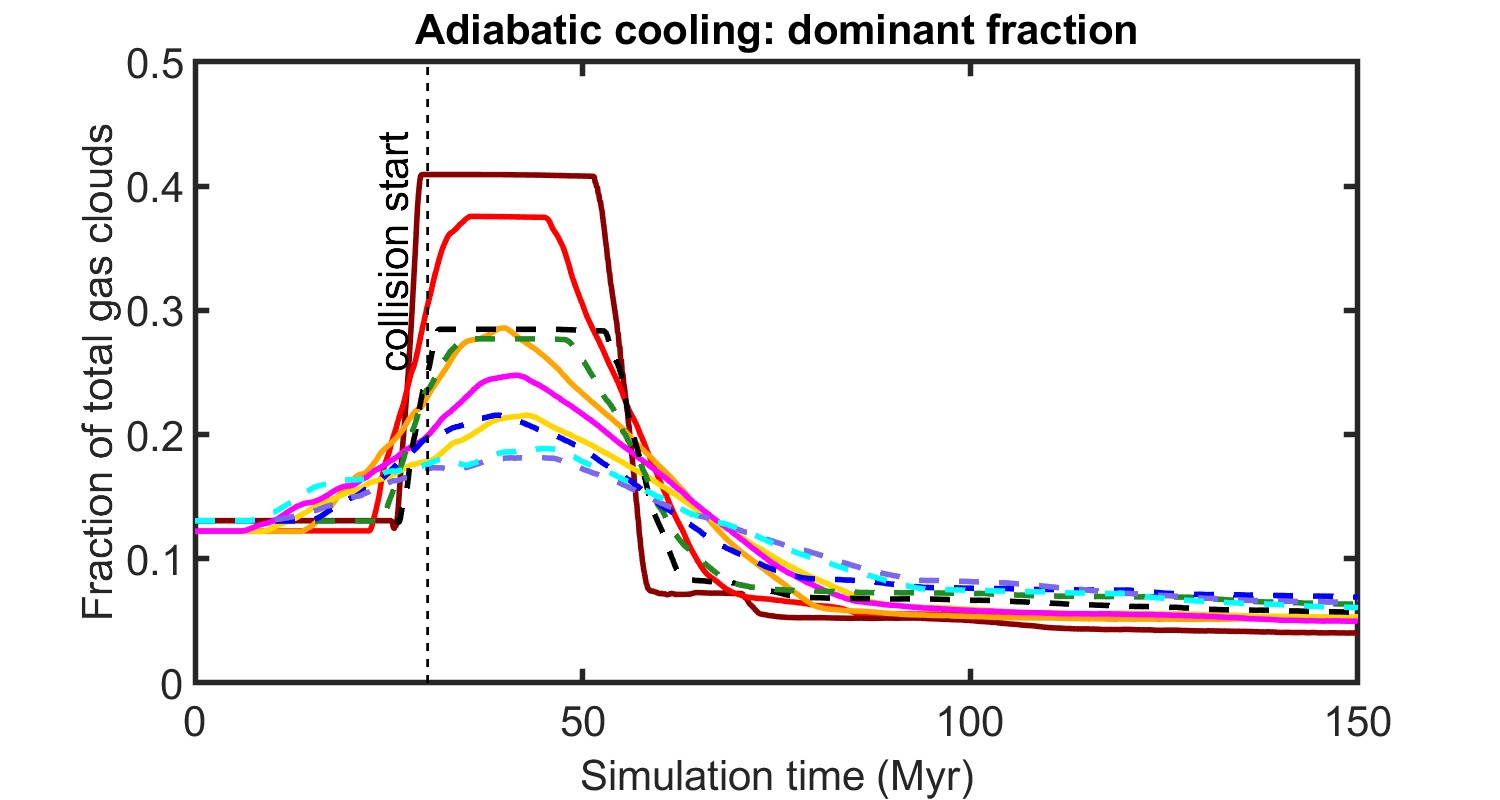}
    } \\
    \subfigure[]{%
      \includegraphics[width=0.39\textwidth]{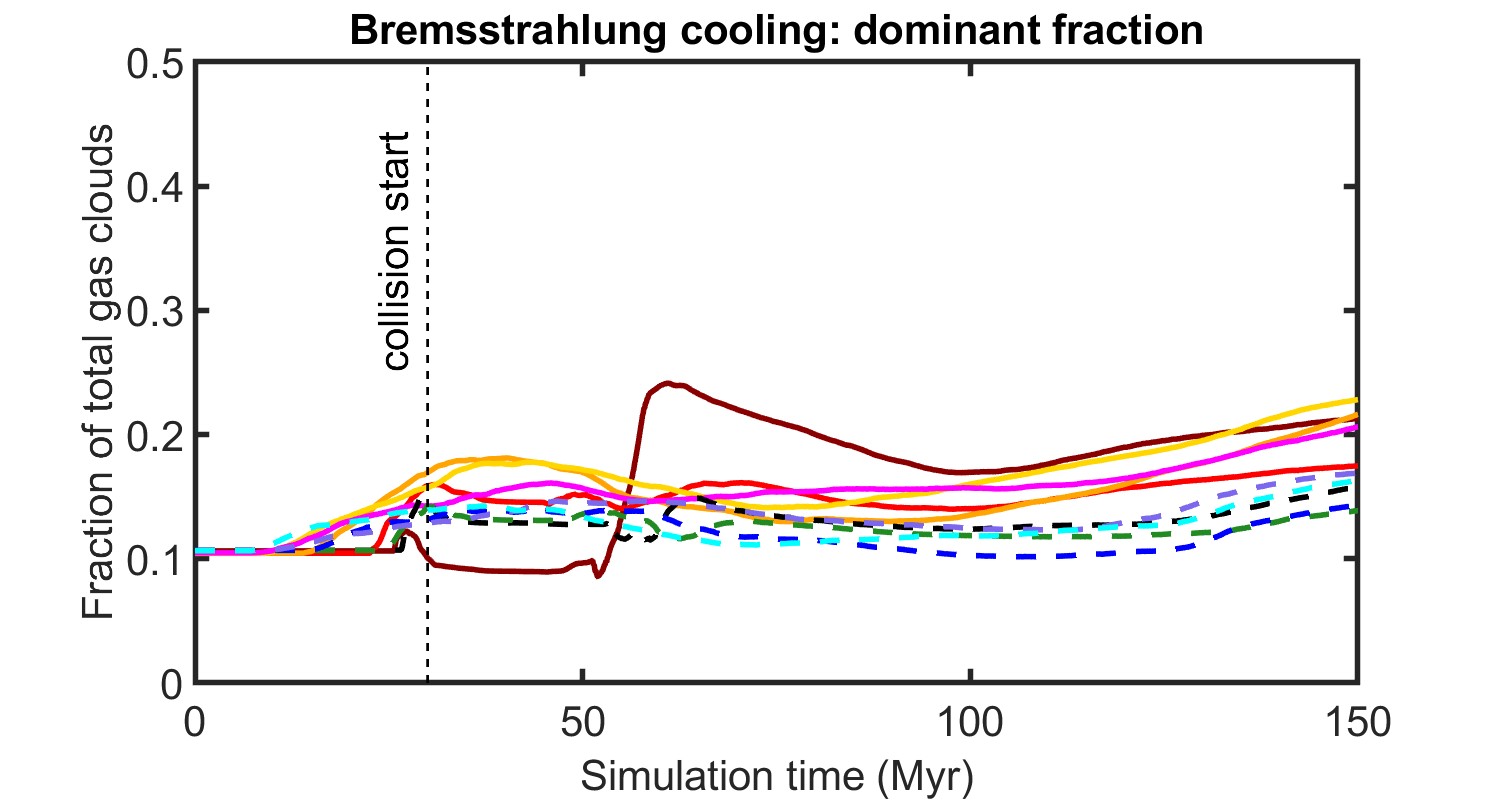}
    } \\
    \subfigure[]{%
      \includegraphics[width=0.39\textwidth]{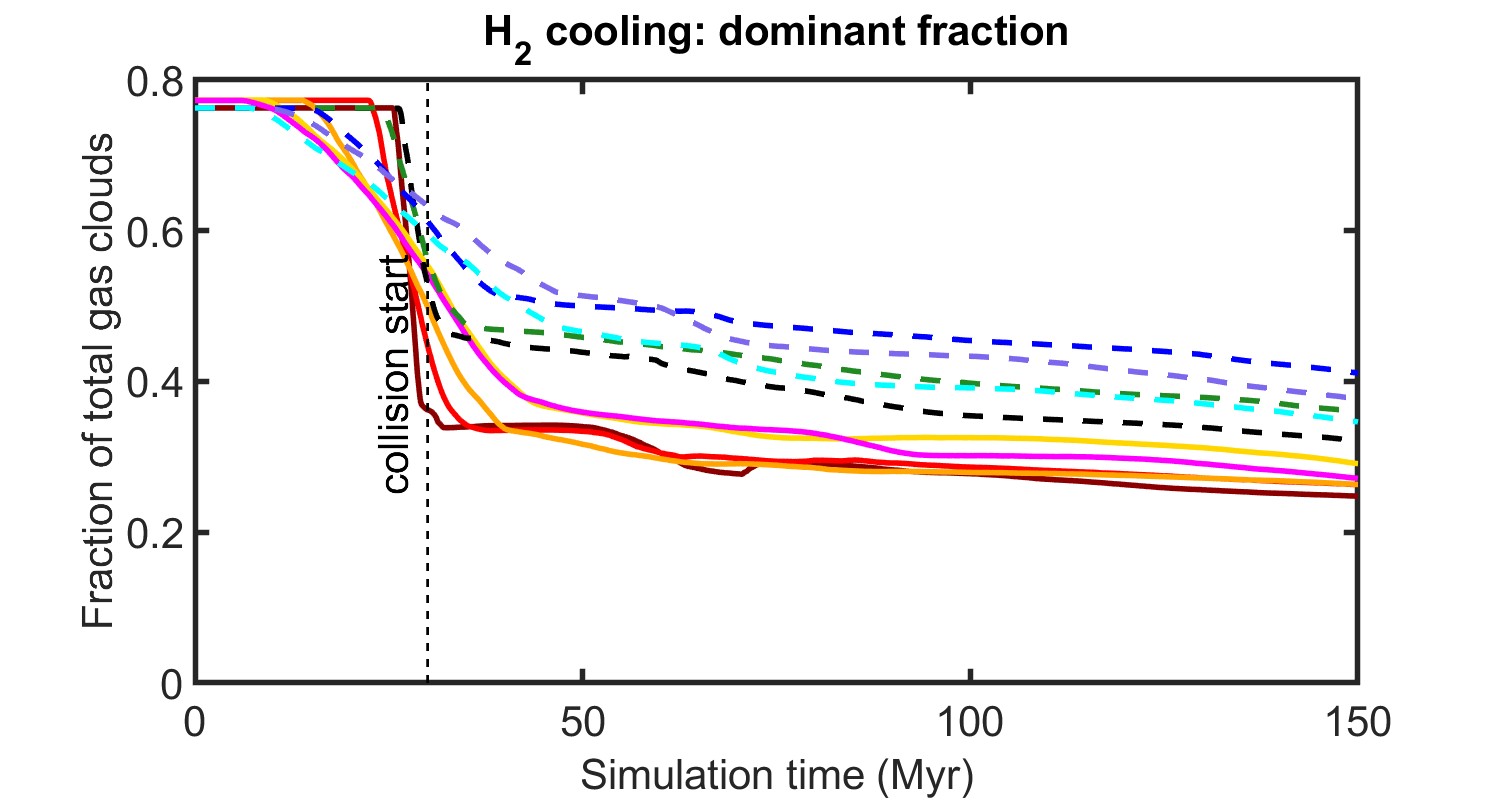}
    } \\
    \subfigure[]{%
      \includegraphics[width=0.39\textwidth]{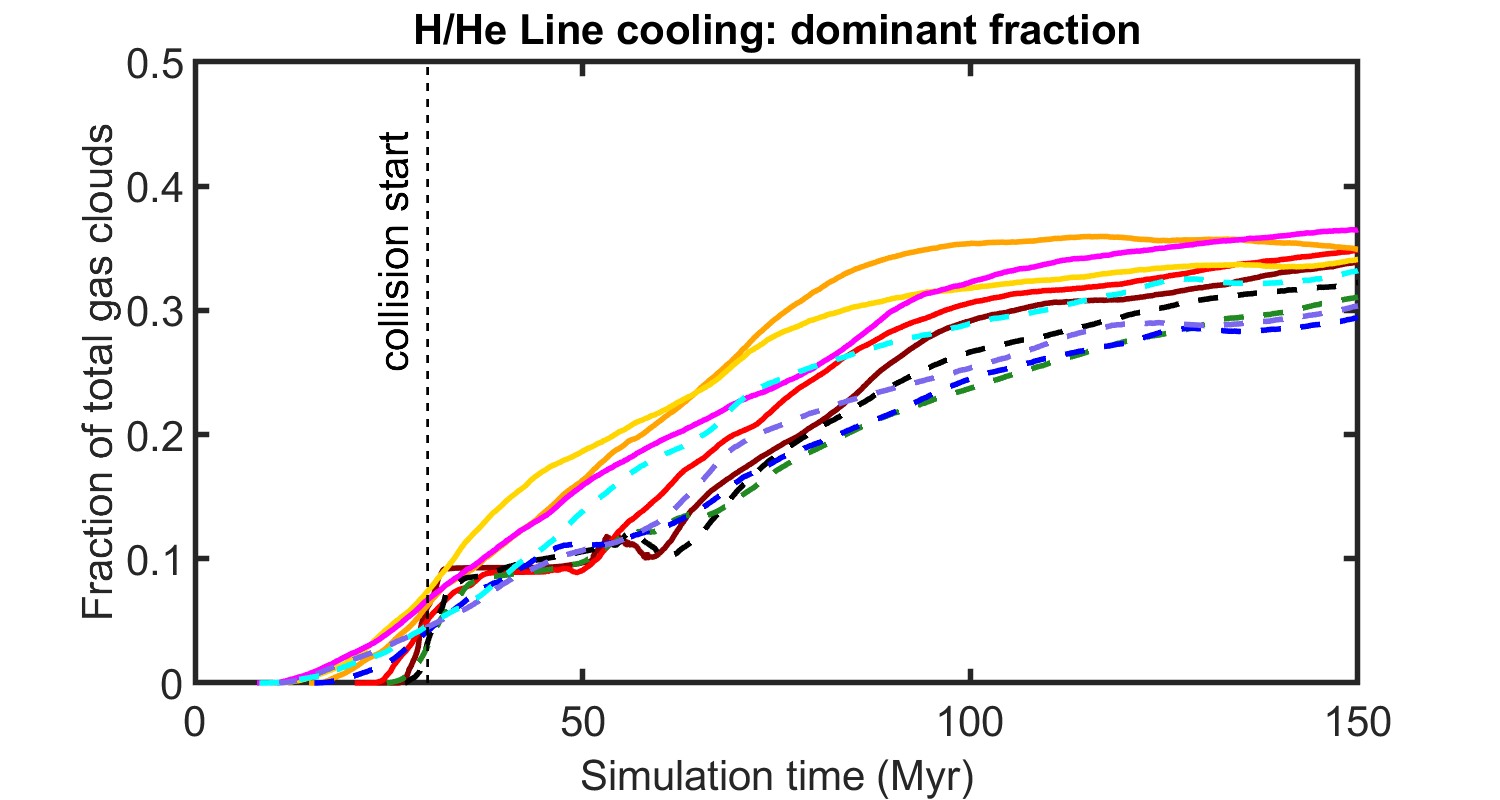}
    } \\
    \caption{Dominance of various cooling pathways for collisions with different offsets and inclinations. (\textbf{a}) The line colors and styles represent different collision types: solid lines indicate collisions with offsets of 500 pc, while dashed lines indicate collisions with offsets of 10 kpc. Each subplot shows the fraction of gas clouds from G1 and G2 where a specific cooling mechanism (\textbf{b}) Adiabatic expansion, (\textbf{c}) Bremsstrahlung, (\textbf{d}) H$_2$, and (\textbf{e}) H/He Line, is stronger than all other cooling mechanisms in the cloud.}
    \label{fig:cooling_dominance}
  \end{center}
\end{figure}

\Cref{fig:cooling_dominance} highlights the relative importance of each gas cooling process in our simulations over time. Panel (b-e) show the fraction of total gas clouds currently dominated by adiabatic cooling, Bremsstrahlung, \hh, and H/He cooling dominance. `Dominance' is when the rate of cooling due to adiabatic expansion exceeds the sum of all other modeled cooling processes. For our model, this means, that the rate of adiabatic cooling is greater than the cooling rate of Bremsstrahlung/H/He/$H_{2}$/Silicon/Oxygen lines, and dust grain cooling. In the legend panel (a) the line colors are used to identify the results from the 10 different disk-disk collisions. A dashed line is used for disk-disk collisions with a 10 kiloparsec offset and solid lines are used for 500 pc offset collisions.

\begin{figure}
  \begin{center}
    \subfigure[]{%
      \includegraphics[width=0.5\textwidth]{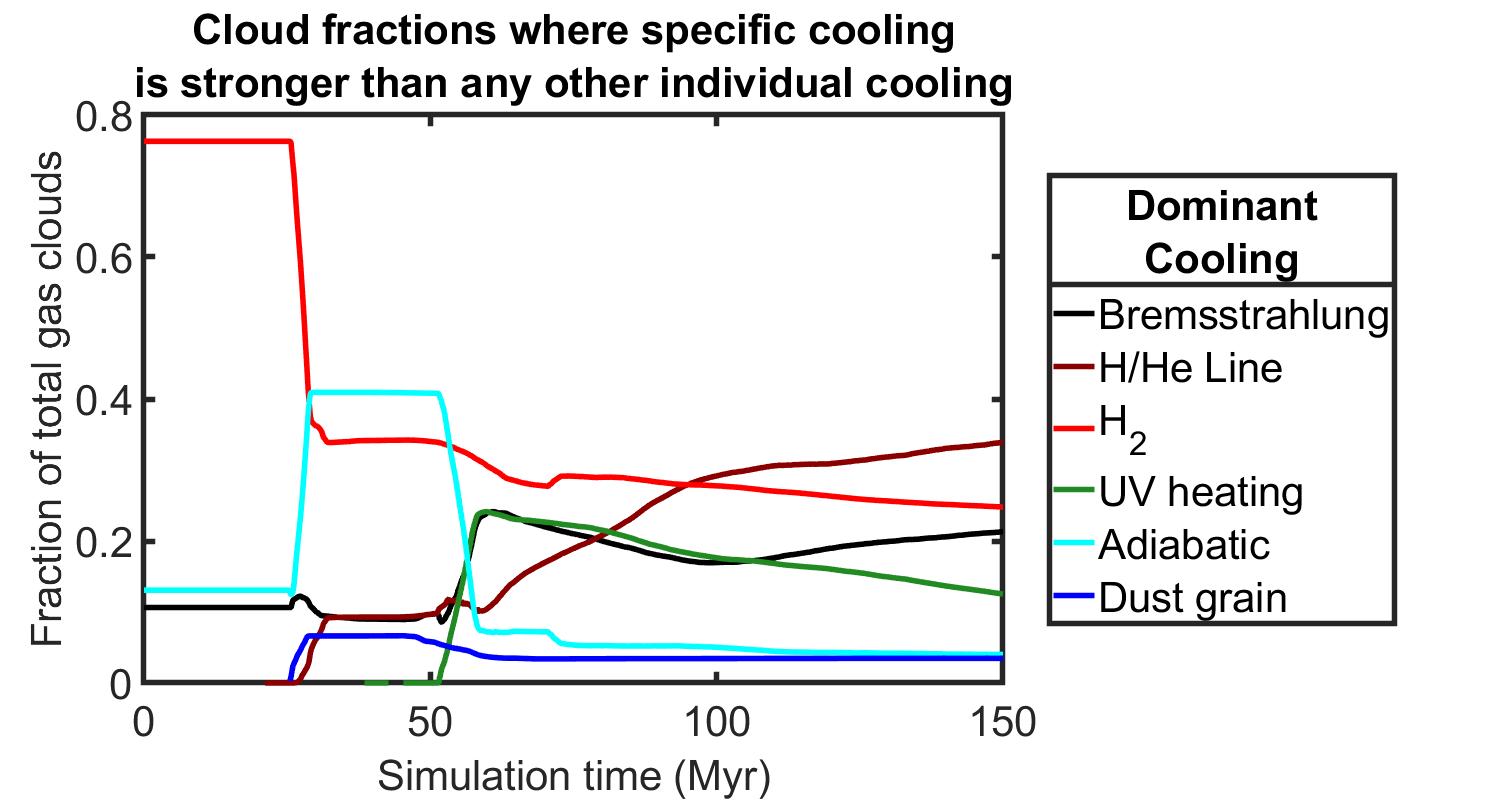}
    } \
    \subfigure[]{%
      \includegraphics[width=0.5\textwidth]{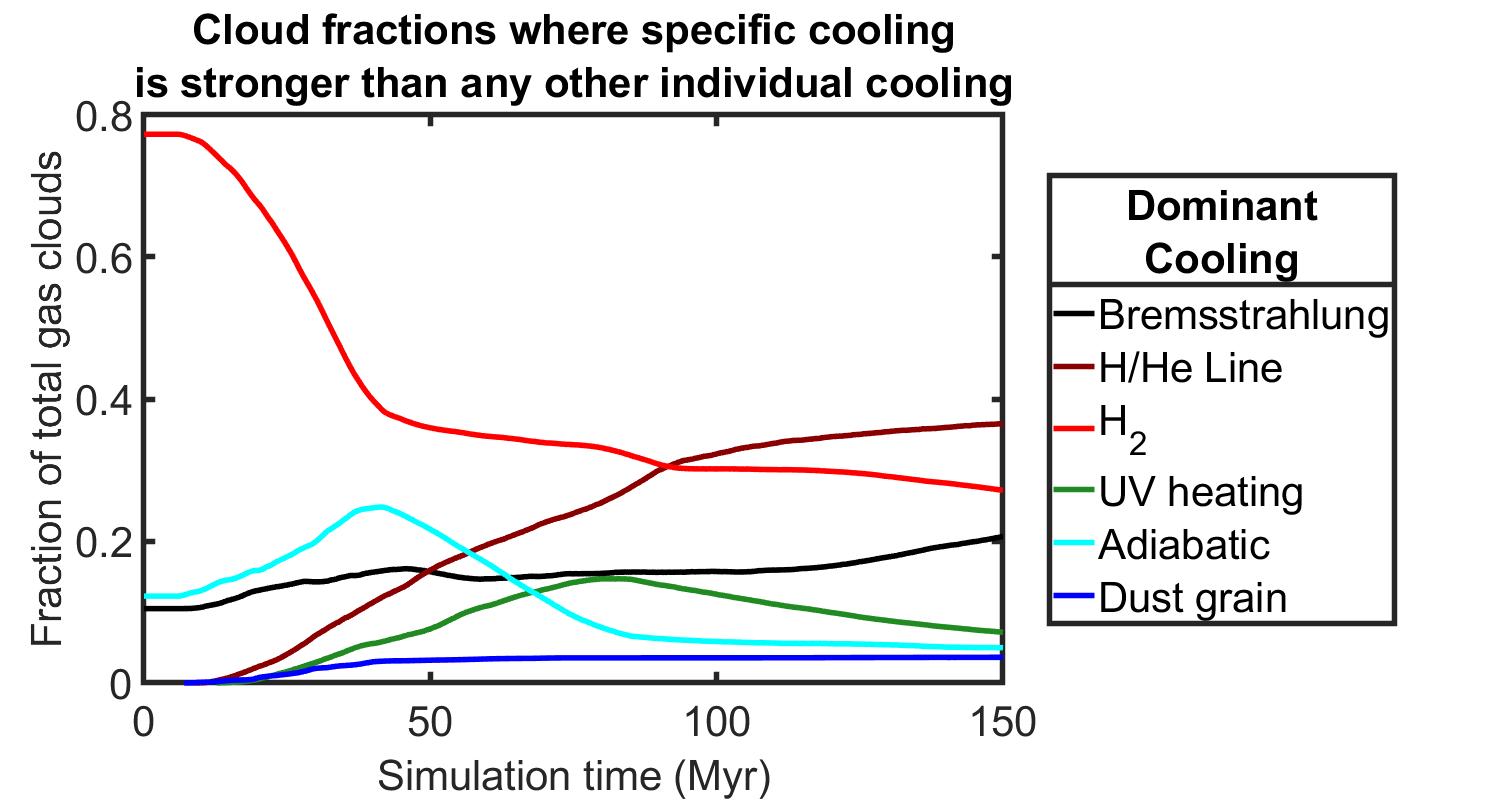}
    }
    \caption{Comparison of the fraction of clouds dominated by any cooling pathway for two 500 pc offset collisions. (a) has a relative inclination of 0\degree{} and (b) is 90\degree{}. Some forms of cooling, such as oxygen, silicon, and CII, are not included since they are never found to be dominant in the simulations.}
    \label{fig:dominant_cooling_in_specific_runs}
  \end{center}
\end{figure}

Initially, \Cref{fig:dominant_cooling_in_specific_runs} reveals that about $13\%$ of all gas particles have temperatures and densities such that adiabatic cooling is strongest. This peaks at 20\% to 40\% of the gas fraction depending on collision type. In all modeled collisions this fraction rises, and later in all collisions, the fraction settles to lower than $10\%$. Adiabatic cooling is most dominant immediately after the disk-disk collision with a decreasing relevance during the reaccretion after 30 Myr.

Low offset-low inclination collisions of counter-rotating disks result in the highest velocity shocks in the gas of both disk galaxies. There appears to be a correlation between the strength of the initial shock waves and the peak fraction of gas clouds with strong adiabatic expansion. 

The fraction of gas clouds dominated by adiabatic expansion decreases as the collision offset increases. Less of the gas in high-offset disk collisions is shocked since the high offset results in smaller parts of the two disk galaxies directly colliding. 

Looking again at the 500 pc offset 0\degree{} inclination collision, there is a well-defined ~30 Myr period where adiabatic expansion is the dominant cooling process. There is a sharp rise in the adiabatic cooling gas fraction immediately after the low offset-low inclination disk-disk collision, which occurs roughly at a simulation time of 25 Myr. Within one mega-year the gas fraction dominated by adiabatic expansion rises to just over $40\%$ and quickly levels off, remaining constant for 30 Myr, before again sharply falling to a gas fraction of under $10\%$. This phenomenon is also visible, to a lesser degree, in all the other models. Presumably, this is because the expansion of the most strongly shock-heated gas is halted by the external medium at the end of this time interval. 

In most instances where adiabatic expansion is the strongest source of cooling, it is also dominant, as defined above. This may be important since adiabatic expansion is one of the only major processes that decrease the density and increase the radius of the gas clouds.

\begin{figure}
  \begin{center}
   \subfigure[]{%
   \includegraphics[width=0.45\textwidth]{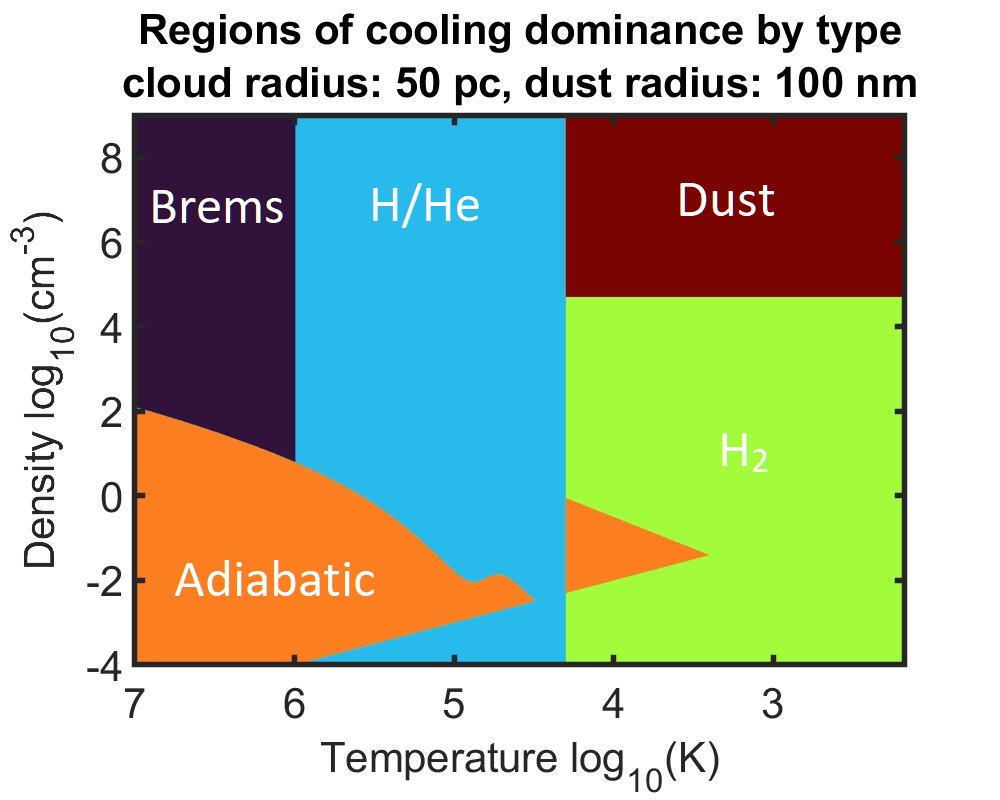}
   } \
   \subfigure[]{%
   \includegraphics[width=0.45\textwidth]{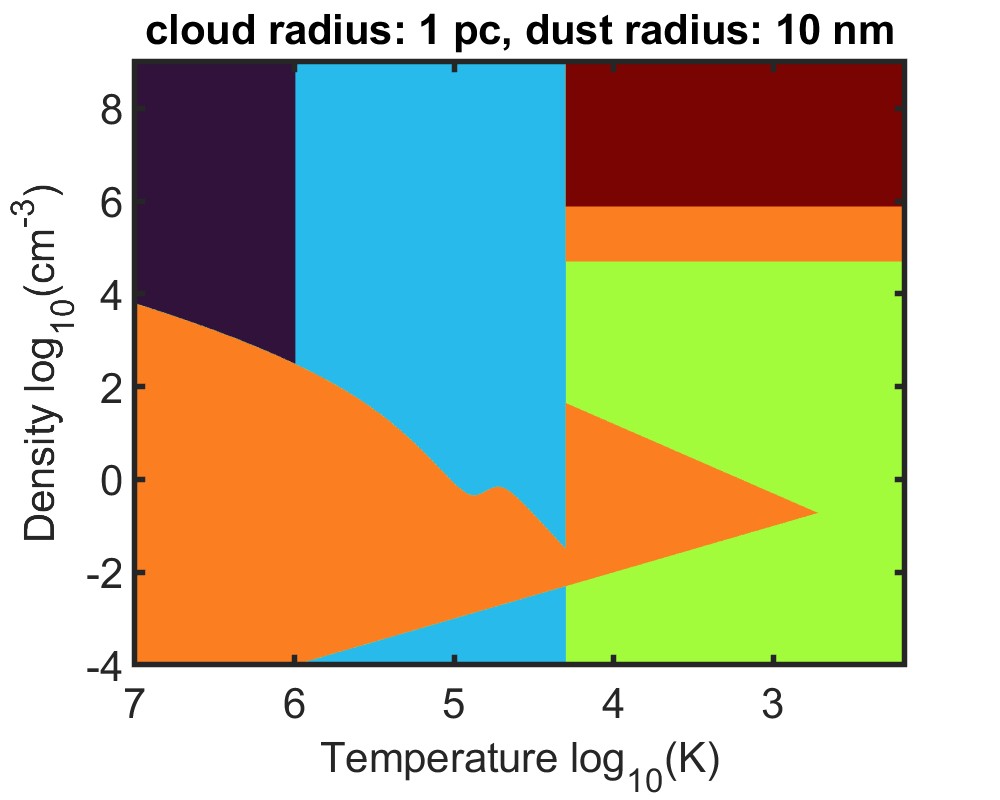}
   } \
   \subfigure[]{%
   \includegraphics[width=0.45\textwidth]{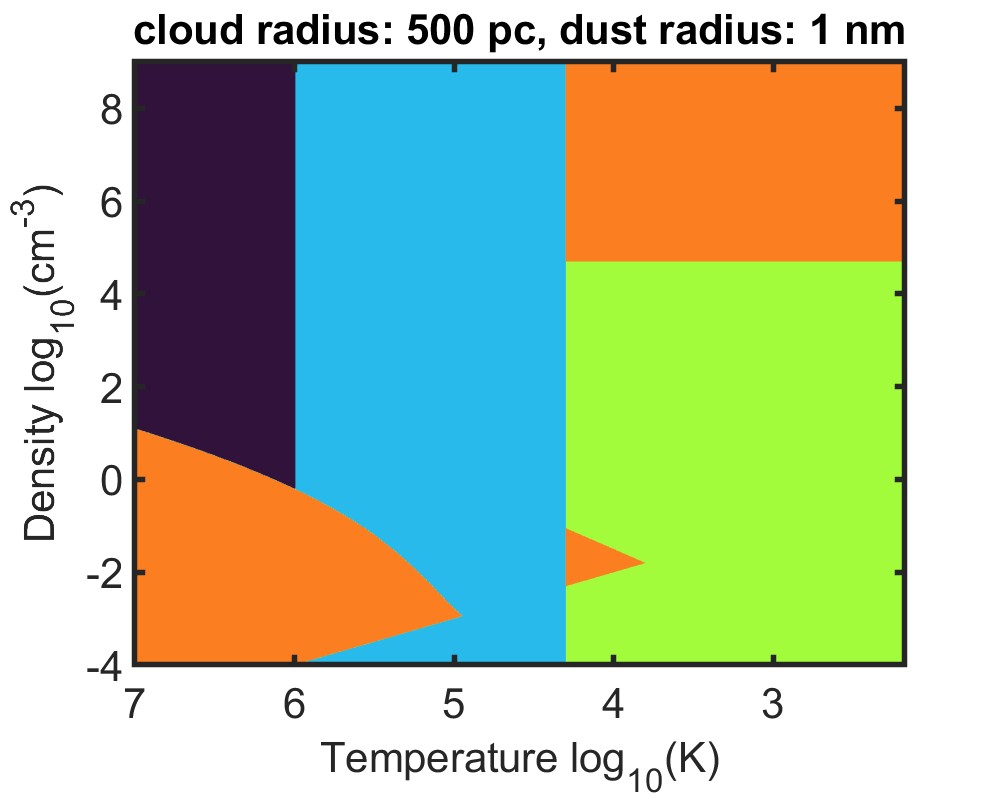}
   }
\caption{Filled contour plots where the colors indicate which type of cooling is dominant in a given temperature and density space. As specified in the title these plots are calculated for a given cloud radius and a uniform dust grain size. A total grain mass equal to $1\%$ the mass of the gas cloud is used for all calculations.} %
\label{fig:regions}
 \end{center}
\end{figure}

\Cref{fig:regions} helps to roughly determine where certain types of cooling are dominant in gas temperature/density parameter space. As a gas cloud grows in radius, the strength of the adiabatic expansion process weakens, reducing the orange regions in the plot. For small gas cloud radii of 1 pc the orange region is large, connects the two regions that are separate in other cases, and reaches to lower temperatures. In the simulation, all gas clouds are initialized with a radius of 50 pc, but with post-shock cooling, the clouds condense and can reach scales of 1 pc. While this cloud condensation increases the density, the reduced temperatures balance that, so there is not sufficient pressure to drive expansion. At temperatures, less than 20,000 K \hh{} cooling remains competitive with adiabatic expansion in most cases.

\subsection{Phase diagrams - gas temperature versus density}

\begin{figure*}
  \begin{center}
   \includegraphics[width=.85\textwidth]{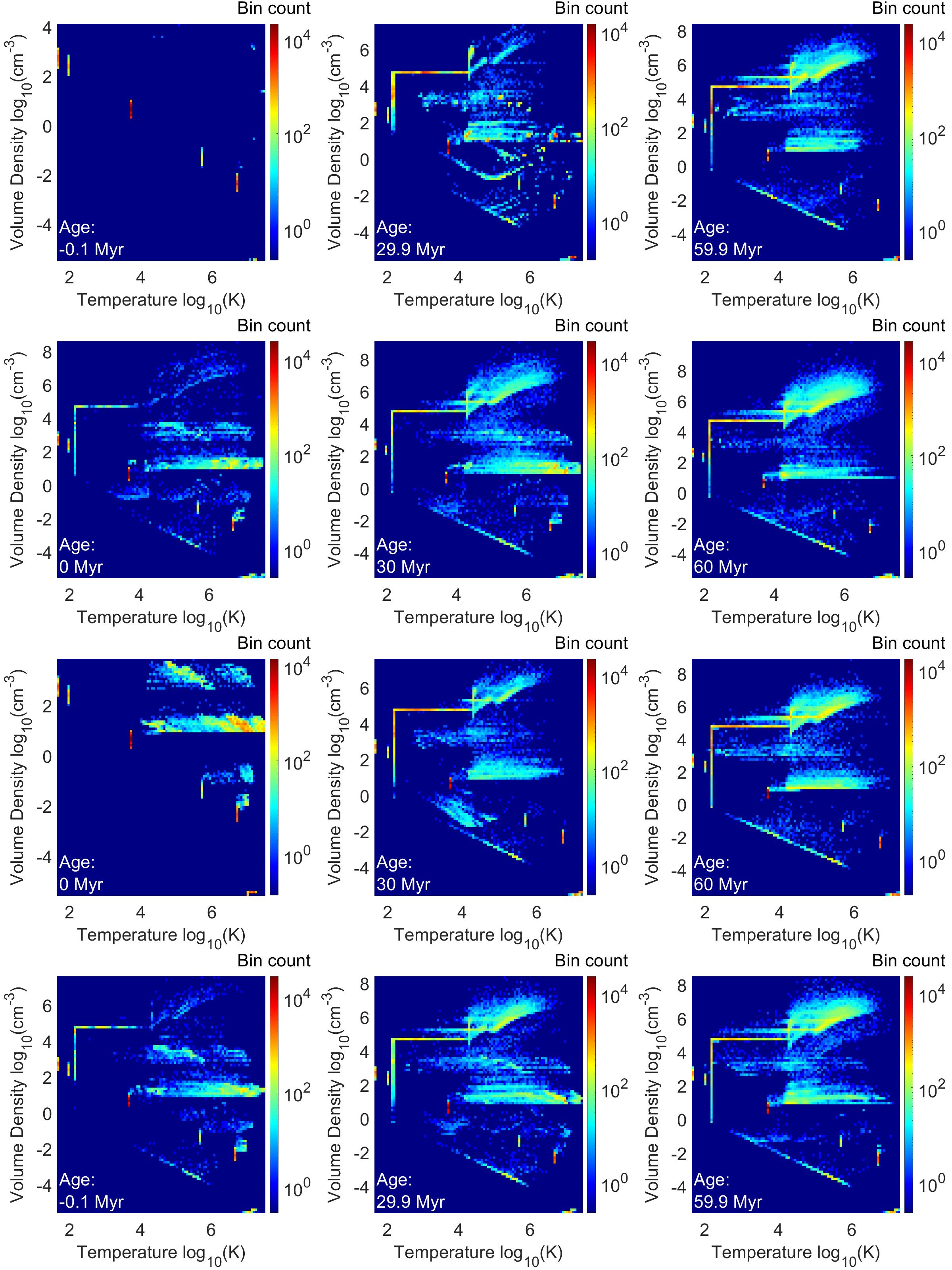}

\caption{Gas temperature vs density phase space diagram. The first row is a zero-degree relative inclination-500 pc offset disk-disk collision. The second row is a ninety-degree relative inclination-500 pc offset disk-disk collision. The third row is a zero-degree relative inclination-10 kiloparsec offset disk-disk collision. The fourth row is a ninety-degree relative inclination-10 kiloparsec offset disk-disk collision. The columns are approximately 0, 30, and 60 Myr after the disk-disk collision, respectively. Gas particles are binned into a 2d-histogram according to the current gas cloud temperature and density. The x-axis, y-axis, and color map are on a logarithmic scale and the color indicates the number of gas clouds per 2d-bin. On the y-axis is the gas volume density per centimeters cubed and the x-axis is the gas temperature, which ranges from 150 to 10 million in Kelvin. There are 108,221 gas clouds in total. Compare these plots with \Cref{fig:regions} to help understand which cooling processes are dominant in each region of temperature and density space.} 
\label{fig:tempvsdensity}
 \end{center}
\end{figure*}

\Cref{fig:tempvsdensity} shows gas temperature versus gas density plots to illustrate the distribution of ISM across the entire interacting system. These sub plots should be referenced to \Cref{fig:regions} to reveal the dominance cooling process occurring within the gas clouds. Each row of \Cref{fig:tempvsdensity} is for a different collision model. The first row is for a 500 pc offset with 0\degree{} inclination, the second row is for a 500 pc offset with 90\degree{} inclination, the third row is for a 10 kiloparsec offset with 0\degree{} inclination and the fourth row is for a 10 kiloparsec offset with 90\degree{} inclination. The columns show three different simulation times, 0, 30, and 60 Myr after the disk-disk collision.

In both high inclination collisions and the high offset-low inclination collision we see populations of shocked gas begin to develop before the disk-disk collision has been completed. In the low inclination, low offset collision shocks occur nearly simultaneously. Before the disks collide all of the ISM in each galaxy is in one of our five assigned gas phases. These are seen as bands in the top left panel of \Cref{fig:tempvsdensity} at 5 discrete temperatures of 50, 500, 5000, \SI{e5}, \SI{e6} K. The density spread is partially the result of the radial exponential profile in the disks and the corresponding scaling of gas cloud densities. 

The thermal evolution in these plots is the direct result of the specific shock or cooling physics in our model. Adiabatically shocked gas moves far to the right (shock-heating) and slightly upward (shocked-cloud-compression) in these plots. Shocked gas fills horizontal regions spread over a temperature range spanning several orders of magnitude. As the gas cools it either moves down and to the left (expansion cooling) or up and to the left. Gas clouds of sufficiently low density are dominated by adiabatic cooling, resulting in a slowly decreasing temperature and decreasing density. These clouds drift down to the left in these plots until they reach a minimum pressure equal to the average pressure of the Intergalactic Medium (IGM), e.g., \citet{Nicastro2018}. This minimum pressure is generally reached after a few tens of millions of years and is seen as a negatively sloped line near the bottom of the temperature-density plots. Once a gas cloud reaches this pressure boundary, other cooling processes may slowly increase the density of the cloud while further reducing its temperature, down to our simulation limit of 150 K. This means an adiabatically cooling cloud would first travel down and to the left before eventually then slowly drifting along the 100 \dunit{} K pressure boundary up and to the left. 

The vertical line that forms far left on the plots is made up of completely cooled (150 K) gas clouds. The flat-thin-horizontal boundary seen at a gas density of \SI{5e5} \dunit{} is where clouds whose cooling is dominated by line cooling reach a critical density where all modeled line cooling is suppressed by particle-particle collisions. A density of \SI{5e5} \dunit{} is the critical density for \hh{} \citet{Roussel_2007}. Above temperatures of 20,000 K, gas cloud densities higher than \SI{5e5} \dunit{} are reached. At such temperatures hydrogen and helium line cooling processes, and at temperatures of millions of degrees, Bremsstrahlung cooling can drive gas cloud number densities up to about \SI{e6} \dunit. These plots provide further evidence that repeated shocks in cloud-cloud collisions can maintain some material at temperatures of 6,000 - 10,000 K (ionized-hydrogen gas) in the splash bridges, on a timescale of several tens of millions of years.

The differences from model to model are tied to the maximum strengths of the shocks. Low inclination and low offset disk collisions produce the highest shock strengths in the case of counter-rotating disks. The counter-rotation of disks with low relative inclinations contributes approximately 500 \kms{} more relative velocity to the cloud-cloud collisions in the disk-disk impacts compared to high inclination cases. The number of gas clouds which reach number densities greater than \SI{5e5} \dunit{} is strongly dependent on the maximum shock strength. The evolution of the temperature and density distribution occurs relatively quickly in low-inclination collisions, as opposed to the prolonged process in high-inclination collisions.

Over time, the initial ISM phases strongly mix in all models. This is seen in the gas temperature versus density plots as a `fuzzing' of the regions where gas clouds gather. Shortly after the collision, shock heating leads to a spread of gas temperatures over a range of \SI{e4} K to \SI{e6} K. Later, the gas clouds spread through density space. Cooling processes also further extend the gas temperature bands down below 1000 K.

\subsection{Turbulence caused by repeated collisions between gas clouds}
Observations show the presence of strong turbulence in the Taffy bridge \citep{gao03,2005xmm..prop..169G,2011AAS...21733527A,2012ApJ...757..138K,peterson12,appleton15,2018ApJ...855..141P,almataffy2019,vollmer2021}. Strongly shearing flows, mixing of gas originating in the two disks, and possibly magnetic effects, are likely drivers of this turbulence. In the remainder of this section, we will focus on cloud-cloud collisions, which are probably the most important aspect of the turbulence generated by mixing and shear. 

Although the Taffy bridge does have a large-scale magnetic field - the nickname derives from radio continuum polarization associated with it - magnetic fields probably only play a significant role in the dynamics of a limited range of scales.

Kelvin-Helmholtz instabilities in the shear flow may be somewhat more important \citep{roediger2015stripped}, but may also operate on a limited range of scales. For non-magnetic, shear flows of comparable density in the opposing flows the instability growth timescale is of order $l/v.$ On kiloparsec scales, with shear velocities of order $100$ km s$^{-1}$, where $l$ is the scale of a Kelvin-Helmholz vortex, and $v$ is the shear scale velocity. This timescale is of order $10$ Myr. Magnetic effects could increase this timescale \citep{das2024magnetic}, but as noted, they are not dynamically important on the bridge scale. Density differences between the shearing layers will increase the timescale, as will the bulk stretching of the bridge as the galaxies move apart. Thus, these instabilities are probably also most relevant on sub-kiloparsec scales, with shocks and cloud collisions being the most important process on scales of order $100$ pc or less. Kelvin-Helmholz flows are not well-represented by sticky particle codes, and a conservative hydrodynamics model would be needed to capture their effects.

\begin{figure*}
  \begin{center}
   \subfigure[]{%
   \includegraphics[width=.45\textwidth]{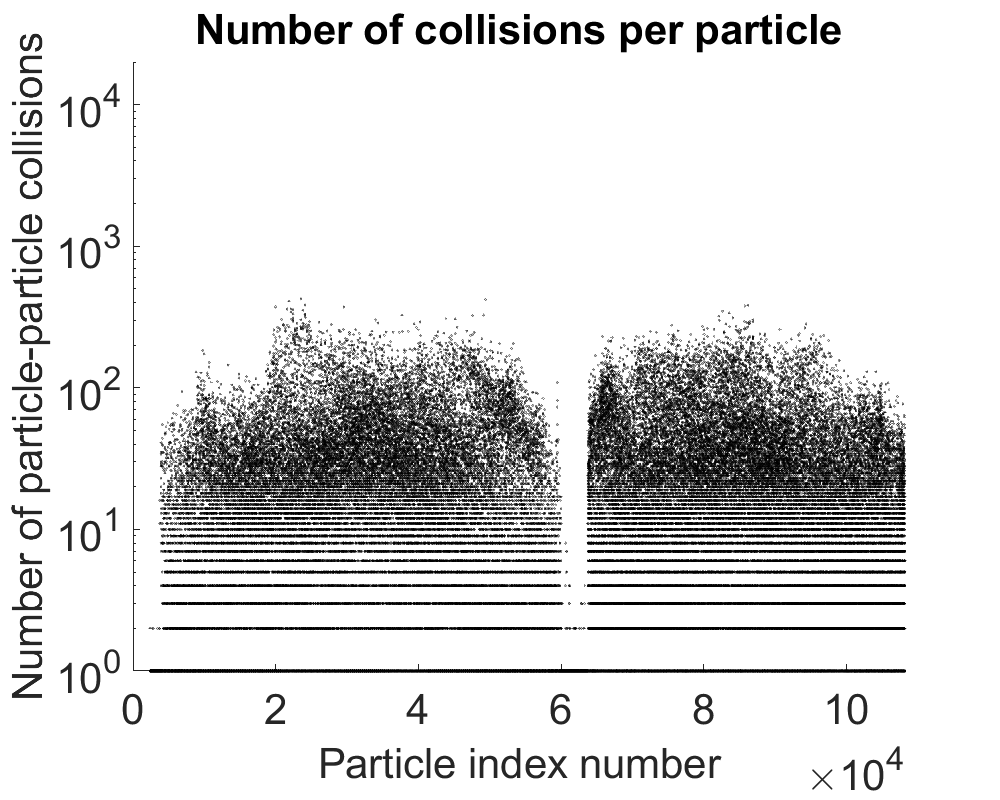}
   }
   \subfigure[]{%
   \includegraphics[width=.45\textwidth]{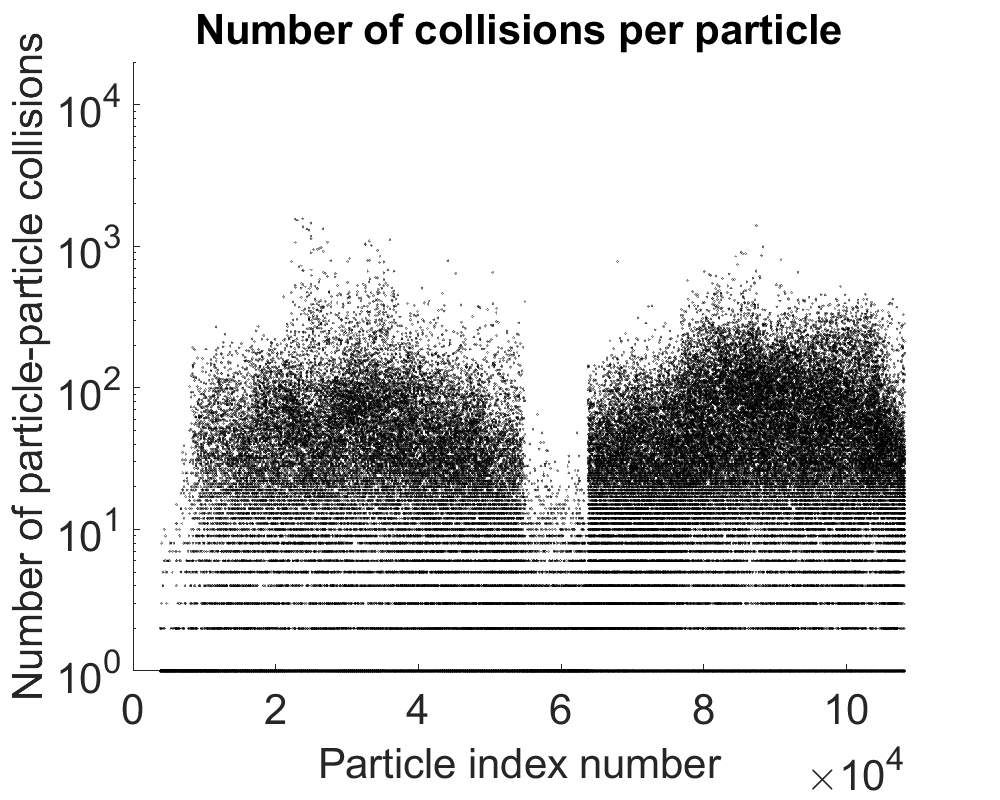}
   }
   \subfigure[]{%
   \includegraphics[width=.45\textwidth]{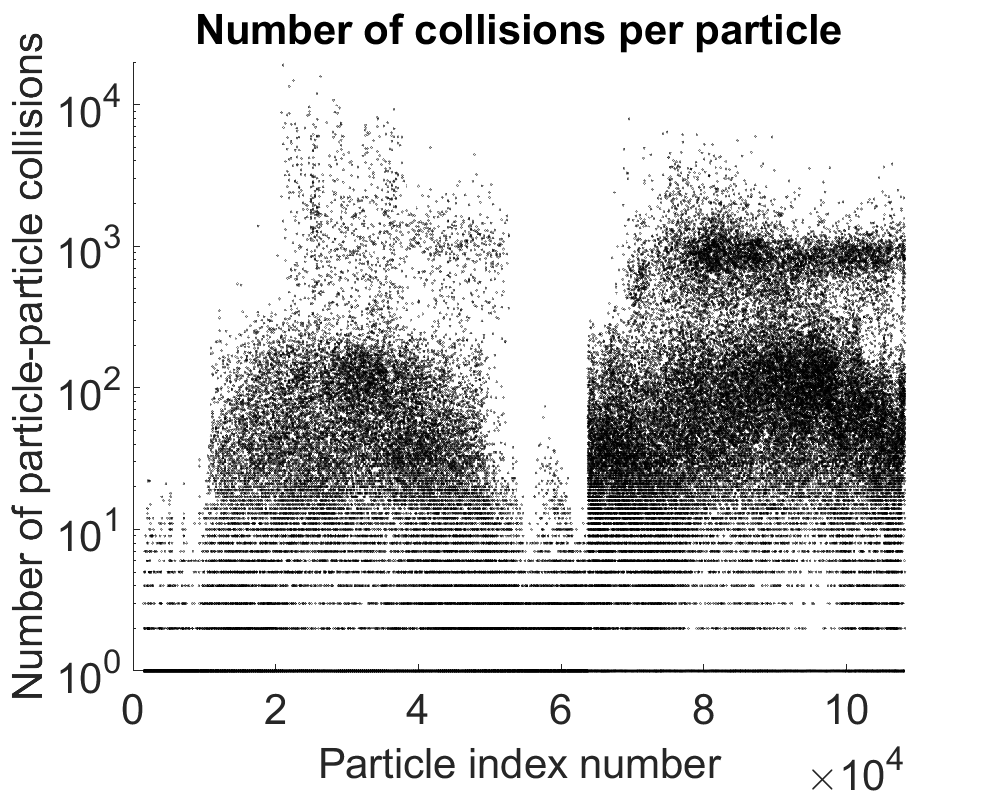}
   }
   \subfigure[]{%
   \includegraphics[width=.45\textwidth]{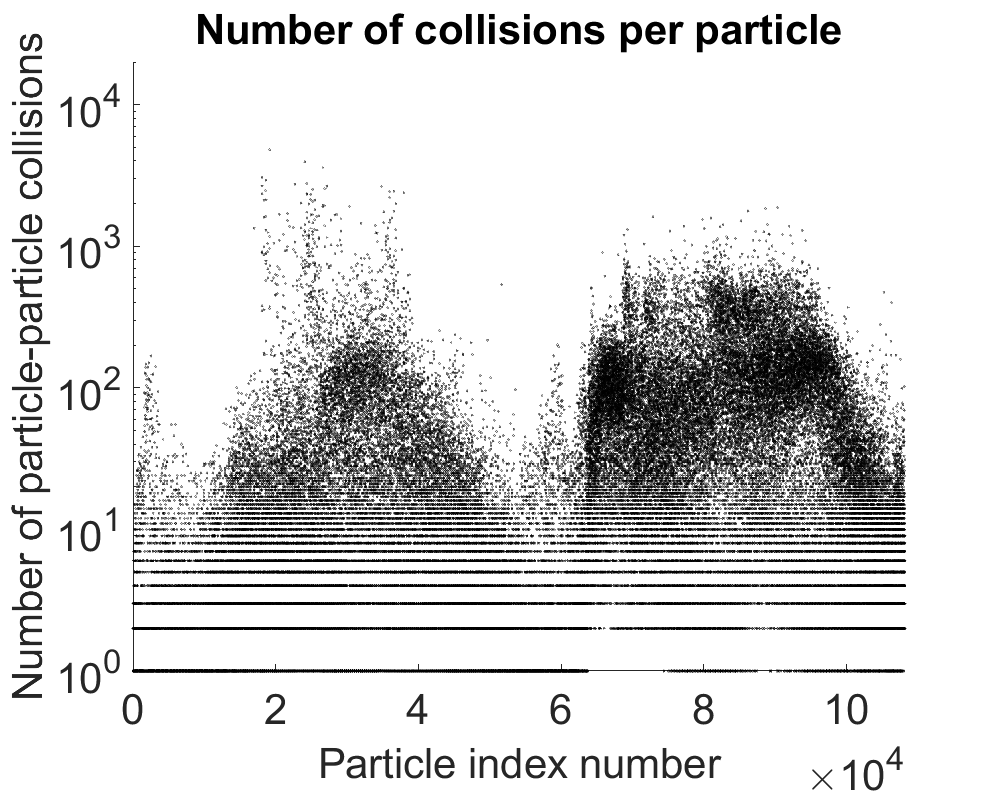}
   }
\caption{Number of collisions per particle for simulation 500 pc offset - (a) 0\degree{} inclination, (b) 45\degree{} inclination, (c) 65\degree{} inclination, (d) 90\degree{} inclination. Particle indexes above about 65,000 are initially in the disk of G2, with the rest originally belonging to G1.} %
\label{fig:numcols}
 \end{center}
\end{figure*}

\Cref{fig:numcols} shows the relationship between the number of collisions per particle and the relative inclination of the disk-disk collision. Specifically, the figure shows 500 pc offset collisions with the following inclinations: panel (a) 0\degree{}, (b) 45\degree{}, (c) 65\degree{}, (d) 90\degree{}. The number of collisions between gas particles increases with inclination up to at least 65\degree{}, but there are fewer collisions per particle in the 90\degree{} models.

Table \ref{table:collision} provides the mean and median values for the time in mega-years between the collisions that all gas particles experience in our 10 models, which probe the collisional inclination and disk offset parameter space for gas-rich disk-disk collisions. These statistics provide further evidence of how the million-degree gas in splash bridges persisted for tens of millions of years. \Cref{fig:abundances}, panel (c) and comparisons of the \textit{extremely hot} gas distributions found in the bottom right sub-panels of \Cref{fig:0inc-rotx0y0z0} shows how the increase in collision inclination supports a significant hot gas phase for tens of millions of years longer than the 0\degree{} inclination collisions. While the lowest inclinations produce the highest hot gas fraction, the lack of a strong cascade of turbulence in the splash bridges of low inclination disk-disk collisions allows that gas fraction to rapidly cool, as illustrated by the steep fall off of the \textit{extremely hot} fraction for low inclinations in \Cref{fig:abundances}. It is perhaps intuitive that high inclination should lead to significantly more turbulence in the splashed gas, as the misalignment of the rotation axes of the two galaxies leads to further collisions as residual angular momentum curves streams of gas into one another. In low inclination collisions, the angular momentum of the galaxy disks remaining after the disk-disk collision carries splash gas in parallel gas streams that interact much less.

\begin{table}
\begin{tabular}{|c|c|c|c|c|c|c|c|c|}
\hline
Offset (kiloparsec) & Incl. (\degree) & Mean & Median\\
\hline
0.5 & 0 & 3.7866 & 0.1\\
0.5 & 20 & 4.0391 & 0.2\\
0.5 & 45 & 2.9497 & 0.1\\
0.5 & 65 & 0.82059 & 0.1\\
0.5 & 90 & 1.8984 & 0.1\\
10 & 0 & 4.8927 & 0.2\\
10 & 20 & 4.2161 & 0.2\\
10 & 45 & 3.1735 & 0.2\\
10 & 65 & 2.6272 & 0.1\\
10 & 90 & 3.0729 & 0.1\\

\hline
\end{tabular}
\caption{The mean and median times, in millions of years. Offset is the collisional offset in kiloparsecs, and inclination, abbreviated Incl., is the relative angle between the colliding gas disks.}
\label{table:collision}
\end{table}

In every model collision, the median time between gas particle collisions is 0.1 Myr, which is the fixed timestep used in the simulation code. This means once a gas cloud-cloud collision occurs, most often it is followed by subsequent collisions in the next step, see Table \ref{table:collision}. The mean time between collisions is much longer, which shows that the particle collision time distribution is not normal. There is a significant fraction of particles experiencing very frequent collisions, presumably in especially turbulent regions.

\section{Results 2: Spectra and Line-of-sight Velocity Maps}
\label{sec:spectra}

\subsection{Plot description}
\label{sec:plot_description}
The double 4-panels of \Cref{fig:0inc-rotx0y0z0,fig:0inc-rotx0y0z90,fig:45inc-rotx0y0z0,fig:45inc-rotx0y0z90,fig:90inc-rotx0y0z0,fig:90inc-rotx0y0z90} show disk-disk collisions at a single relative inclination at a time of 30 Myr. \Cref{fig:0inc-rotx0y0z0,fig:0inc-rotx0y0z90} are for 0\degree{} inclination collisions, \Cref{fig:45inc-rotx0y0z0,fig:45inc-rotx0y0z90} figures are 45\degree{} inclination collisions, and \Cref{fig:90inc-rotx0y0z0,fig:90inc-rotx0y0z90} figures are 90\degree{} inclination collisions. In these figures, the four-panel group labeled (a) illustrates collisions with offsets of 500 pc, and the group labeled (b) shows ten kiloparsec offset collisions, both at the given inclination. The four subpanels of each group are arranged as follows: in the top left is the total \los{} profile, in the top right is a color-coded \los{} plot of all gas particles, the bottom left panel only contains gas particles defined in section \ref{sec:methods} as Critically High Density (\textit{high density}), and the bottom right panel contains only \textit{extremely hot} gas. The color red in the particle \los{} figures corresponds to a redshift (into the page), and blue represents a blue-shift.

The top left panel of \Cref{fig:0inc-rotx0y0z0,fig:0inc-rotx0y0z90,fig:45inc-rotx0y0z0,fig:45inc-rotx0y0z90,fig:90inc-rotx0y0z0,fig:90inc-rotx0y0z90} illustrate the kinematics of line-like velocity profiles of our simulated systems. Line-of-sight velocity will be abbreviated as \los{} in this section. The line-of-sight velocities are placed in bins of 5 \kms{} width and are normalized by the total number of gas particles. In each velocity profile figure, we have plotted the total system gas in black, the \textit{extremely hot} gas in red, \textit{high density} gas in blue, \textit{initially dense gas} gas in green, and ionized-hydrogen gas in magenta. In all spectra figures, the color-line spectra (\textit{extremely hot}, \textit{high density}, \textit{initially dense gas} and ionized-hydrogen) are scaled up by 5x; the y-axis fractions are accurate for the black (total gas) spectra. While the latter four curves may relate to specific observational wavebands, the first (black, total) does not. However, it may be useful for comparisons to future simulations.

The text in the left corner of the spectra panels is the angle of the current point-of-view, the collision offset, relative collision inclination, and the age. The point-of-view (POV) is given as three angles of rotation around the Cartesian X, Y, and Z axes. The disk of G1 always lies within the XY-plane, and the Z axis is perpendicular to this and through the center of G1. G2 is always initially offset along the negative y-axis. Rotations about the z-axis always result in an edge-on view of G1, while a 90\degree{} rotation about the x-axis will show the system face-on. The age of the system is defined as the approximate time since the disk-disk collision. For the highly inclined collisions, the collision continues over a duration of about 10 Myr, so the age is taken from the time when the galaxy centers are at their nearest approach.

\subsection{Plot results}
The time evolution of the \los{} full profile across all of our models tends toward a similar state by about 100 Myr after the disk-disk collision. While the profiles still differ at that time, they all develop a strong \los{} peak around zero \kms. At the time of the disk-disk collision, marked in the first panel as Age: 0 Myr, we see two double peaks in the velocity profile from the two, as yet mostly unaffected, disk galaxies. By 30 Myr, the two pairs of double peaks are completely disrupted, and the \los{} profile is broadened into many peaks. After 100 Myr, \los{} profile range has been reduced to within a range of $\pm$ 200 \kms, centered on 0 \kms. This is a trend across all of our models. The cause is the substantial amount of momentum exchanged between the gas of both disks. The initial disruption in the galaxy collision does not completely remove the highest velocity gas but does spread the gas out over velocity space. The gas at the highest velocities is slowly reduced in a smooth and gradual process on timescales of tens of millions of years. 

\begin{figure}
  \begin{center}
   \subfigure[]{%
   \includegraphics[width=.5\textwidth]{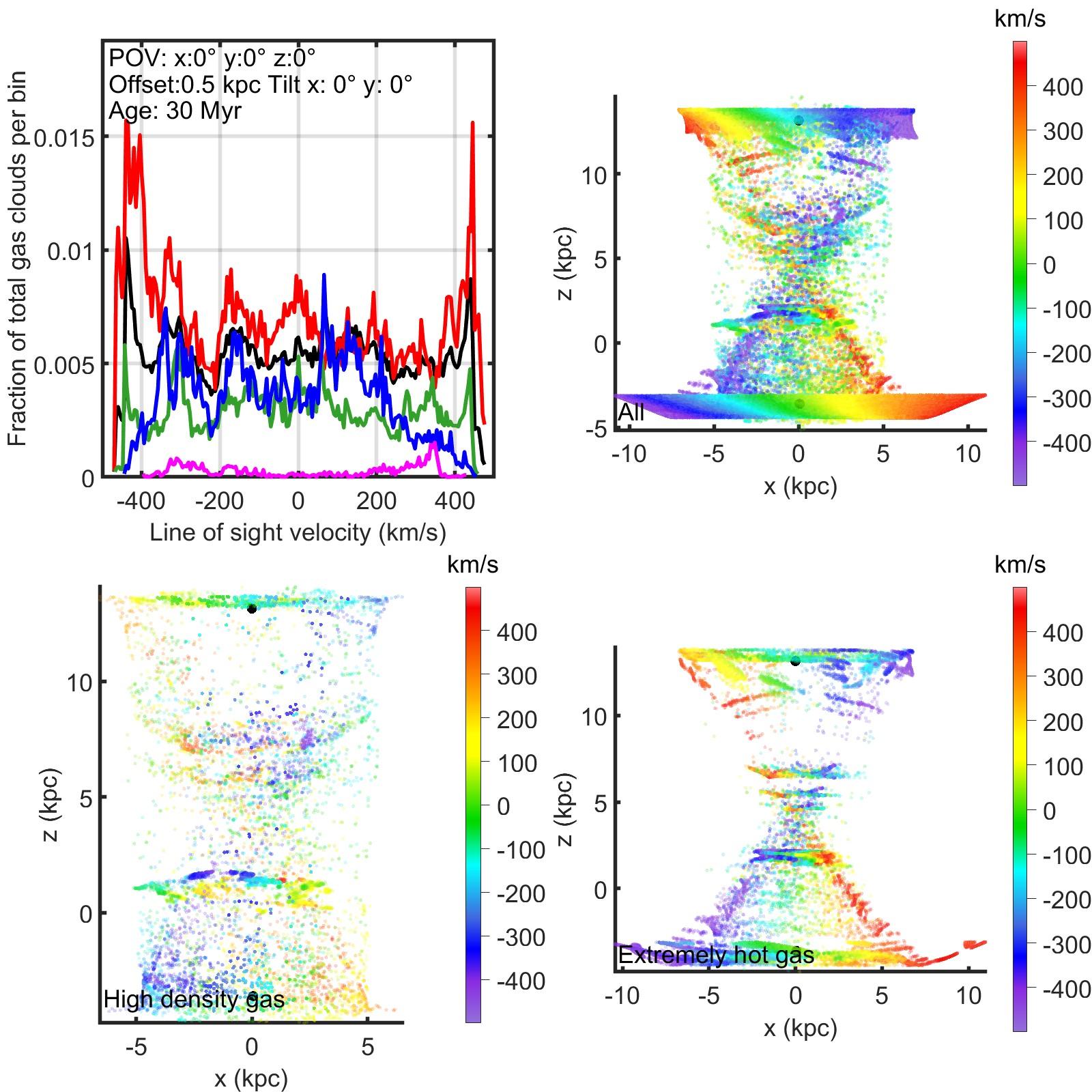}
   }
   \subfigure[]{%
   \includegraphics[width=.5\textwidth]{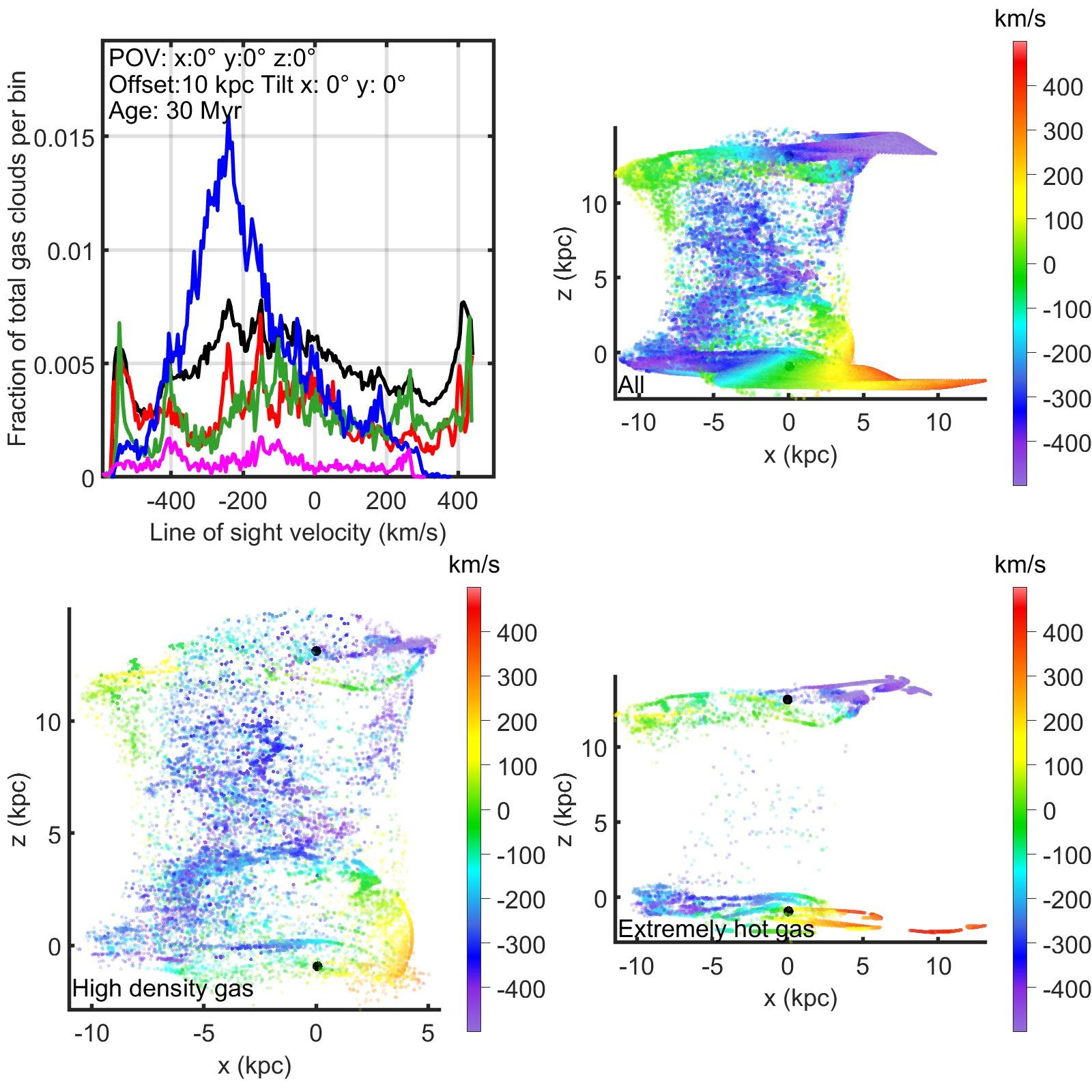}
   }
\caption{Edge-on view of G1. Age: 30 Myr. Collision: 0\degree{} inclination. Panel (a) 500 pc offset. Panel (b) The ten kiloparsec pc offset. Information on each sub-panel can be found in section \ref{sec:plot_description}.} %
\label{fig:0inc-rotx0y0z0}
 \end{center}
\end{figure}

Bridge clouds in the 0\degree{} inclination collisions show a residual rotation about the axis of the parent galaxy. The angular momentum of the gas in the bridge is the net angular momentum resulting from the cloud-cloud collisions in the direct disk-disk collision. This is seen most clearly in part (a) of \Cref{fig:0inc-rotx0y0z0}, where the bridge has an `X' shape with one leg of the X red-shifted, and the other leg of the `X' shifted blue, with a different sense above and below the bridge center. Near the center of the bridge, just below the middle of the `X,' there is a central bridge disk, first observed in the simulations of \citet{yeager19}. This flat disk of gas near the middle of the splash bridge rotates at a rate different from the rotation rates of G1 or G2, again with angular momentum derived from the sum of the particles from both disks. 

\begin{figure}
  \begin{center}
   \subfigure[]{%
   \includegraphics[width=.5\textwidth]{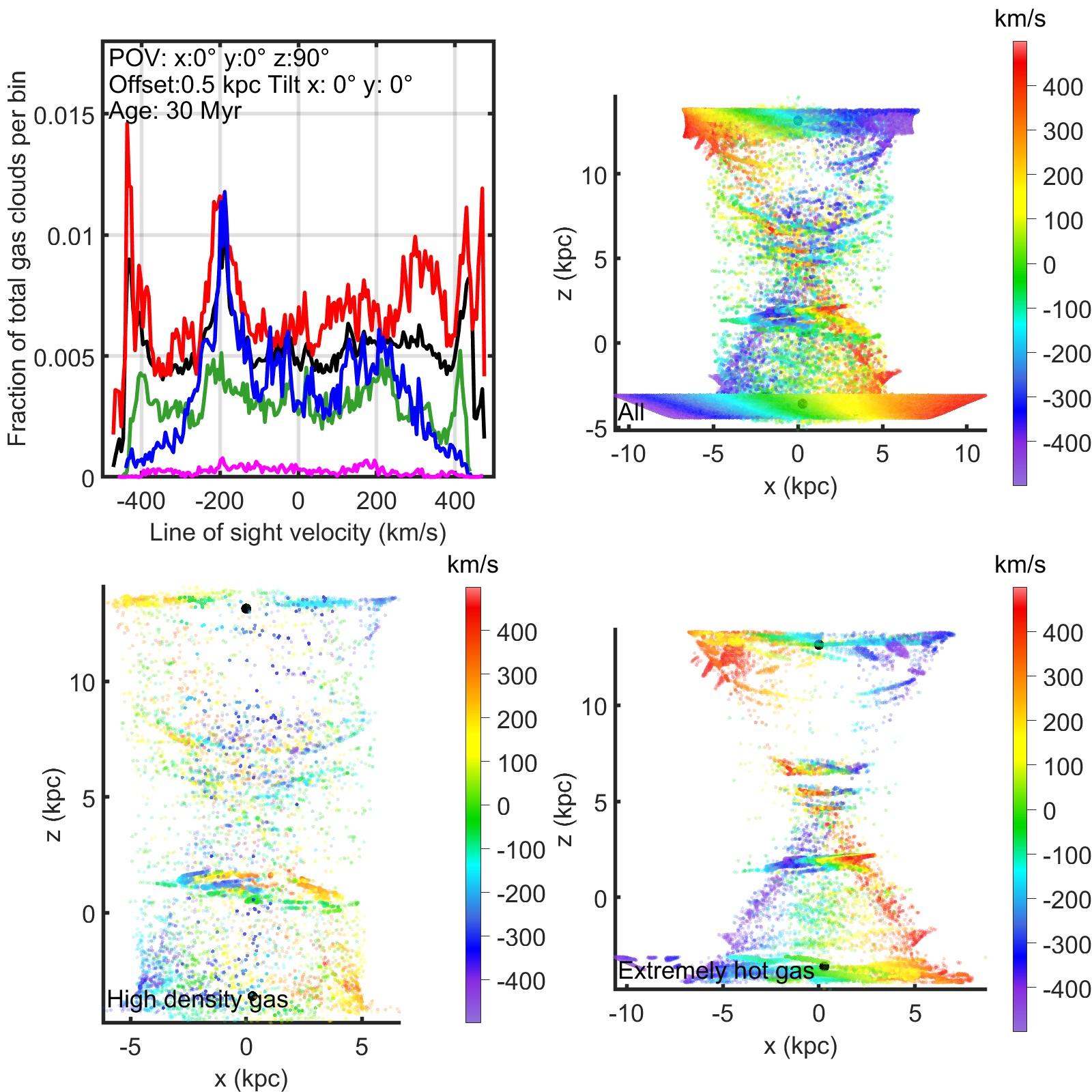}
   }
   \subfigure[]{%
   \includegraphics[width=.5\textwidth]{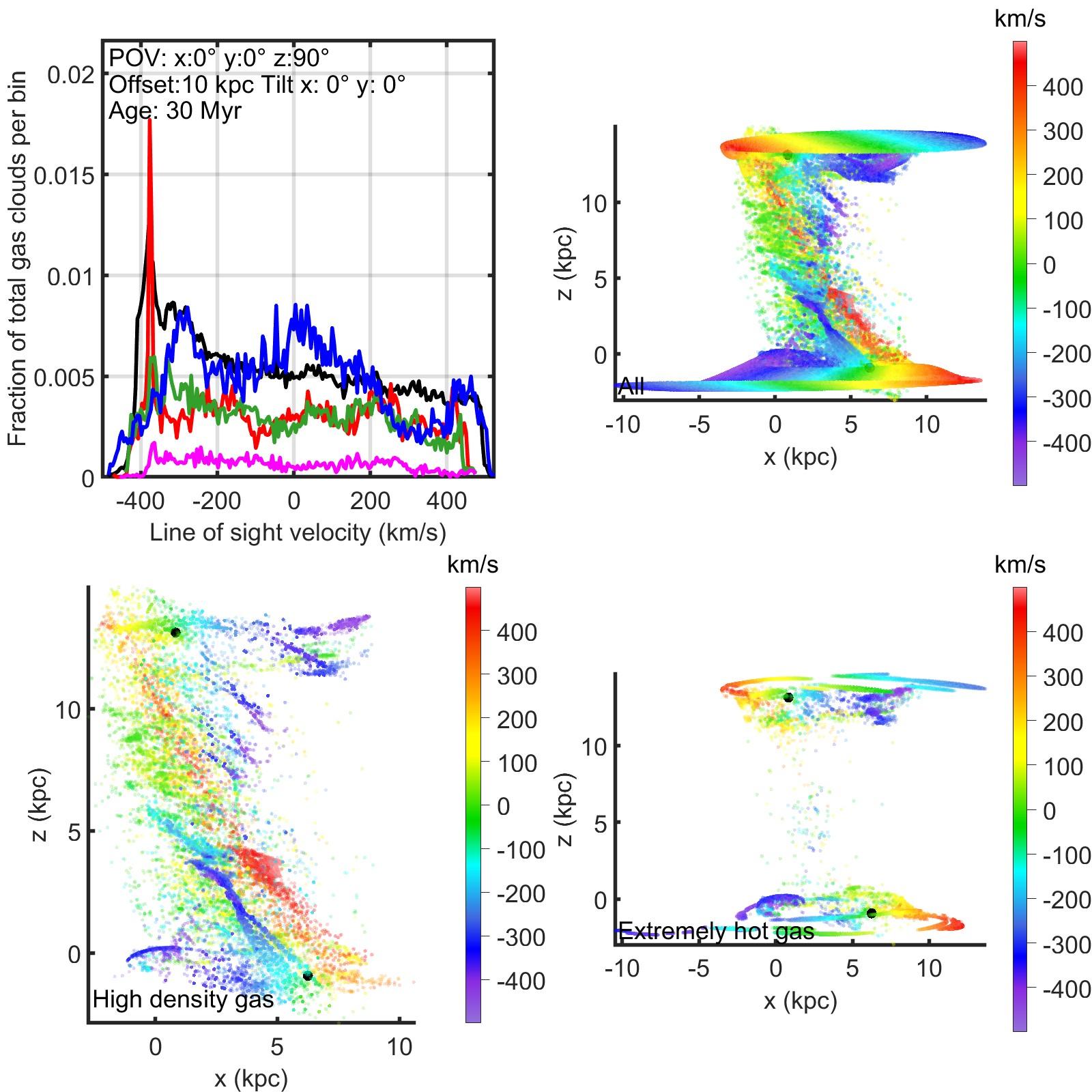}
   }
\caption{Edge-on view of G1, POV: 90\degree{} relative to \Cref{fig:0inc-rotx0y0z0}. Age: 30 Myr. Collision: 0\degree{} inclination. Panel (a) 500 pc offset. Panel (b) The ten kiloparsec pc offset. Information on each sub-panel can be found in section \ref{sec:plot_description}.} %
\label{fig:0inc-rotx0y0z90}
 \end{center}
\end{figure}

\begin{figure}
  \begin{center}
   \subfigure[]{%
   \includegraphics[width=.5\textwidth]{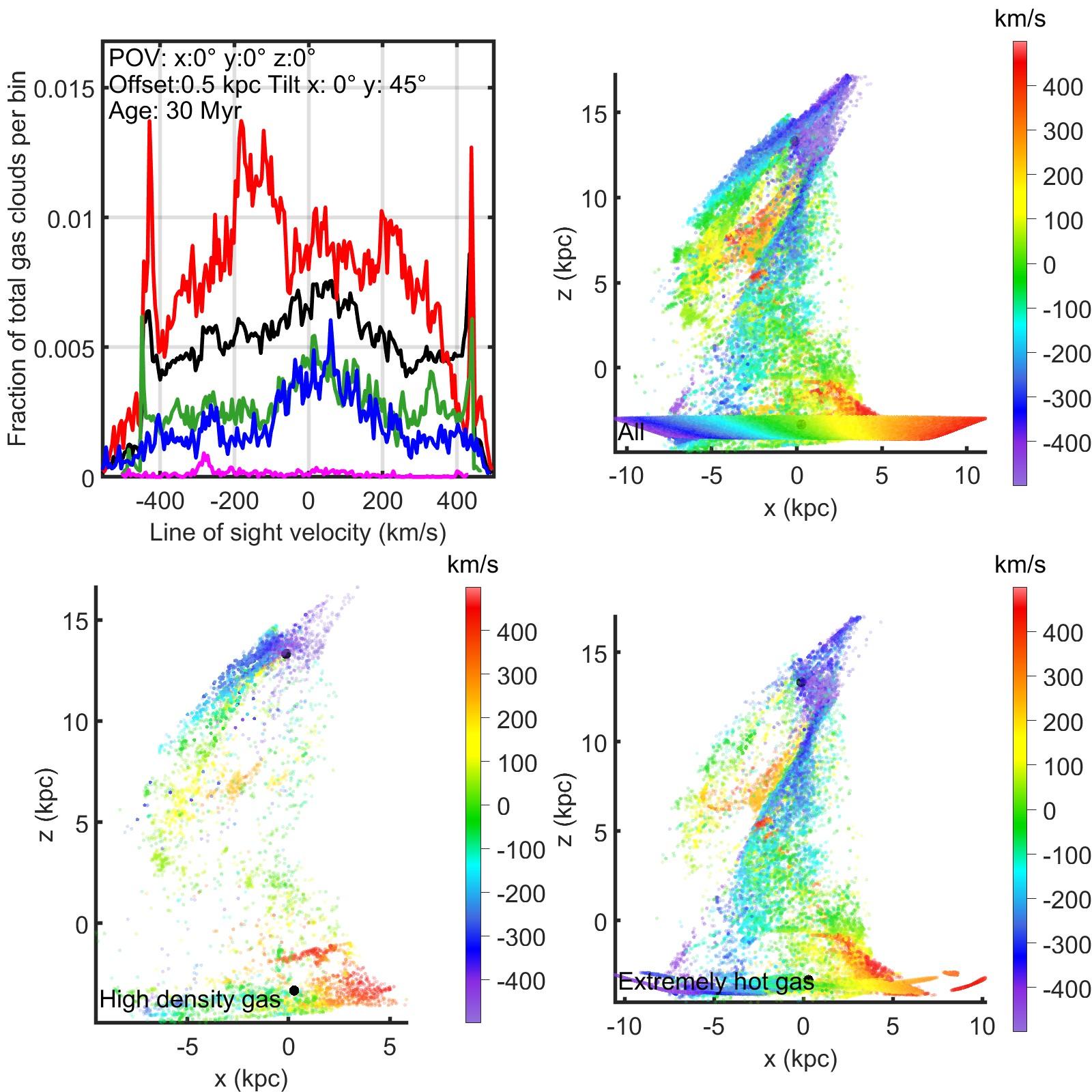}
   }
   \subfigure[]{%
   \includegraphics[width=.5\textwidth]{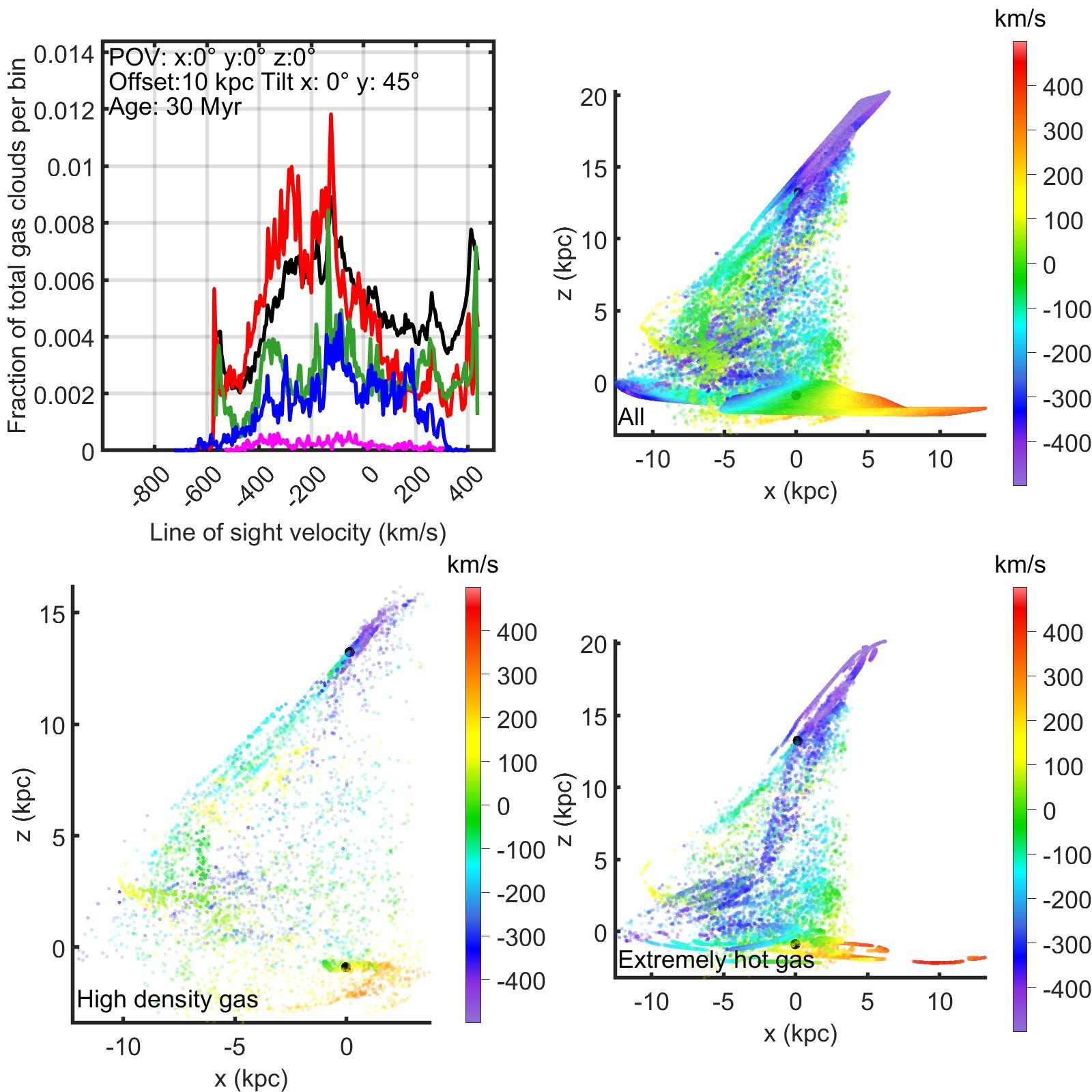}
   }
\caption{Edge-on view of G1. Age: 30 Myr. Collision: 45\degree{} inclination. Panel (a) 500 pc offset. Panel (b) The ten kiloparsec pc offset. Information on each sub-panel can be found in section \ref{sec:plot_description}.} %
\label{fig:45inc-rotx0y0z0}
 \end{center}
\end{figure}

\begin{figure}
  \begin{center}
   \subfigure[]{%
   \includegraphics[width=.5\textwidth]{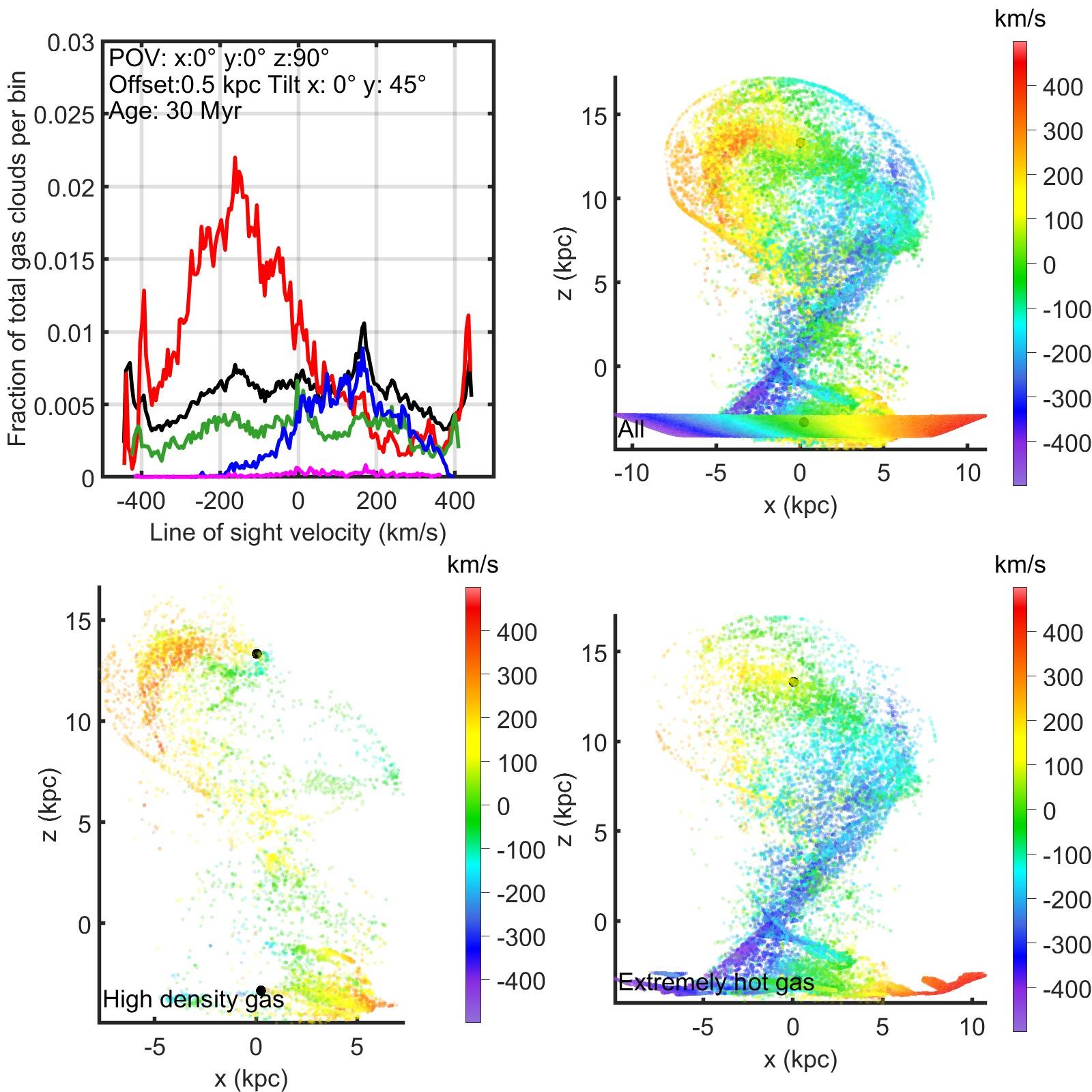}
   }
   \subfigure[]{%
   \includegraphics[width=.5\textwidth]{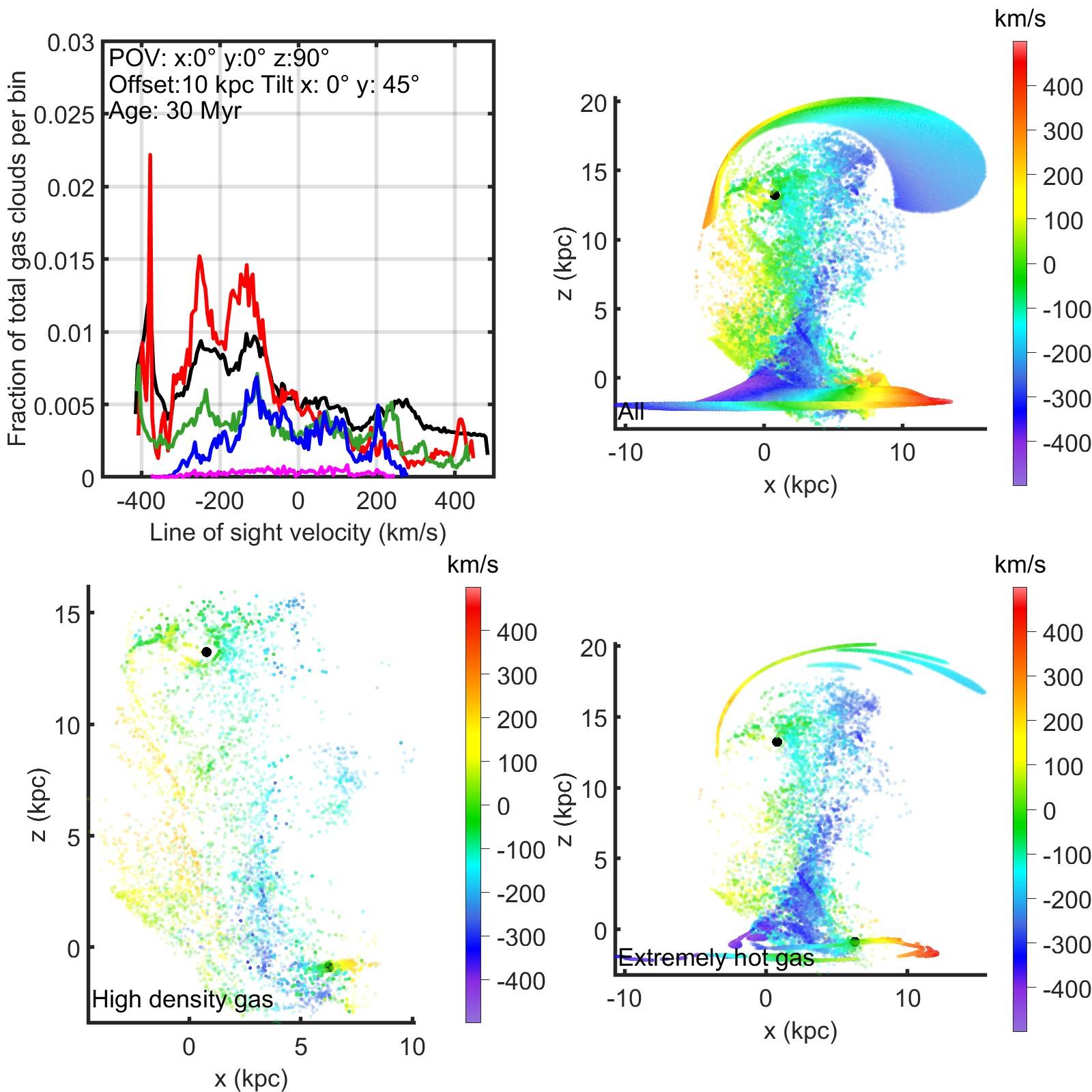}
   }
\caption{Edge-on view of G1, 90\degree{} relative to \Cref{fig:45inc-rotx0y0z0}. Age: 30 Myr. Collision: 45\degree{} inclination. Panel (a) 500 pc offset. Panel (b) The ten kiloparsec pc offset. Information on each sub-panel can be found in section \ref{sec:plot_description}.} %
\label{fig:45inc-rotx0y0z90}
 \end{center}
\end{figure}

The different viewpoints of \Cref{fig:0inc-rotx0y0z90} highlight the strong contrasts in morphology in both the case of the 500 pc offset collisions in panel (a) and the ten kiloparsec offset collisions in panel (b). The \los{} spectra also show different morphologies. \Cref{fig:0inc-rotx0y0z90} is an edge-on view. The 500 pc offset collision in panel (a) has three peaks, at about -400, -200, and 400 \kms{}, and overall the spectra are flat for other velocities between -400 and 400 \kms. In the ten kiloparsec offset collision, there is a peak at -400 \kms{} with a continuous trailing off of the velocities toward 400 \kms. These differences are the same in an off-axis point of view but are different in \Cref{fig:0inc-rotx0y0z0}, which is another edge-on view perpendicular to \Cref{fig:0inc-rotx0y0z90}.

The low offset collision produces substantially more long-lasting (over 30 Myr) hot gas spread over the entire splash bridge than the ten kiloparsec offset collision. This is a result of both more gas in the disk-disk overlap, which also makes repeat cloud collisions more likely, and the fact that the two disks are more aligned, which with the counter-rotation, results in higher initial collision velocities for more gas clouds. The low offset 0\degree{} collision produces hot has found along the bridge `X' structure, or in reality on an `X' rotated around the z-axis producing a double cone region. The cold gas is concentrated around the edge of the double cone region. At this time, gas clouds in the hot region likely must undergo repeated cloud-cloud collisions to maintain the high temperatures. Gas clouds away from this structure have fewer multiple collisions and can cool. A ring of hot gas exists in the central bridge disk where the double cones pass through it, while the regions at larger radii in the central bridge disk contain only cold gas.

When the relative disk inclination is increased, less angular momentum is canceled out in the disk-disk collision. Any semblance of the `X' or double coned region seen in the 0\degree{} collision is washed out by even modest inclination increases.

\begin{figure}
  \begin{center}
   \subfigure[]{%
   \includegraphics[width=.5\textwidth]{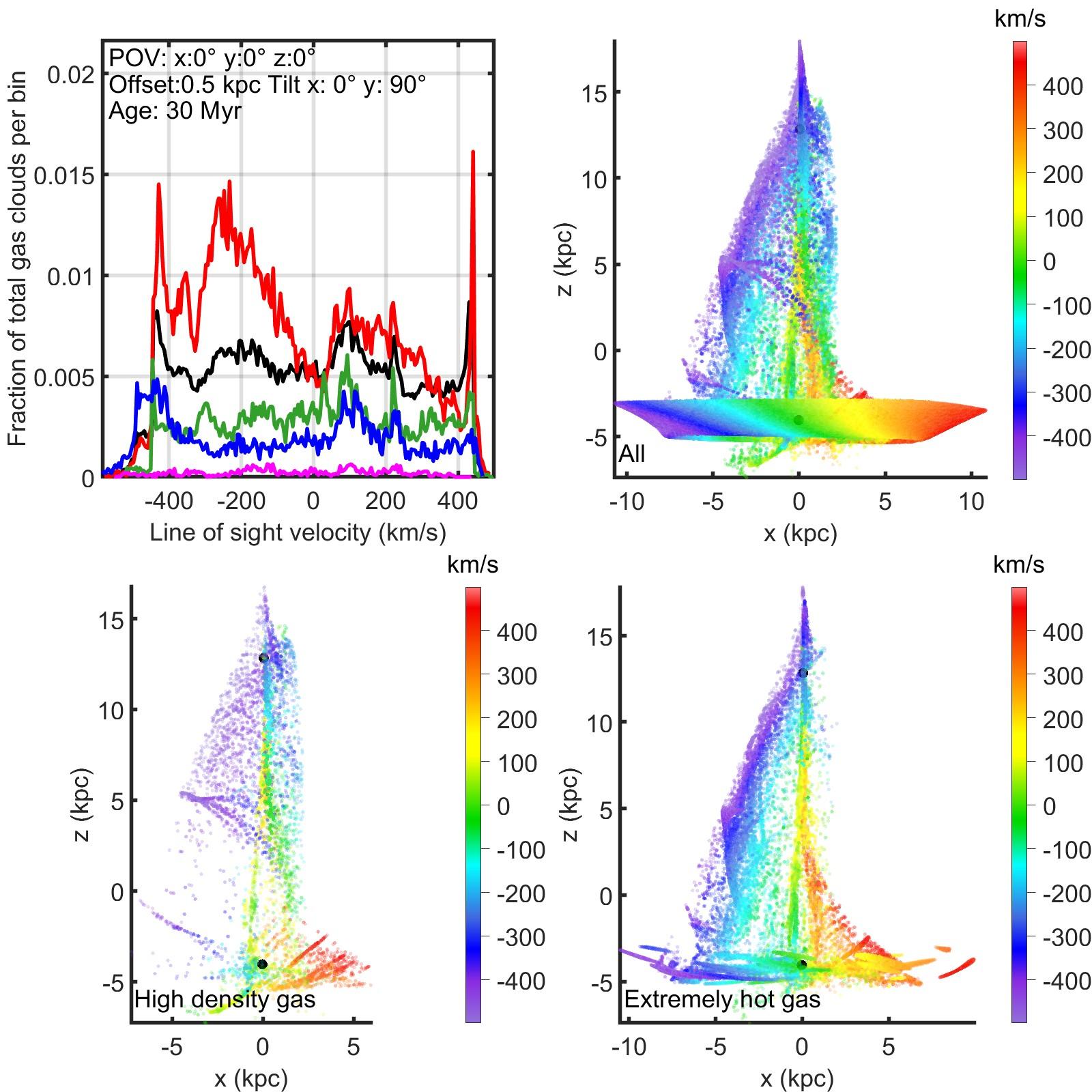}
  ` }
   \subfigure[]{%
   \includegraphics[width=.5\textwidth]{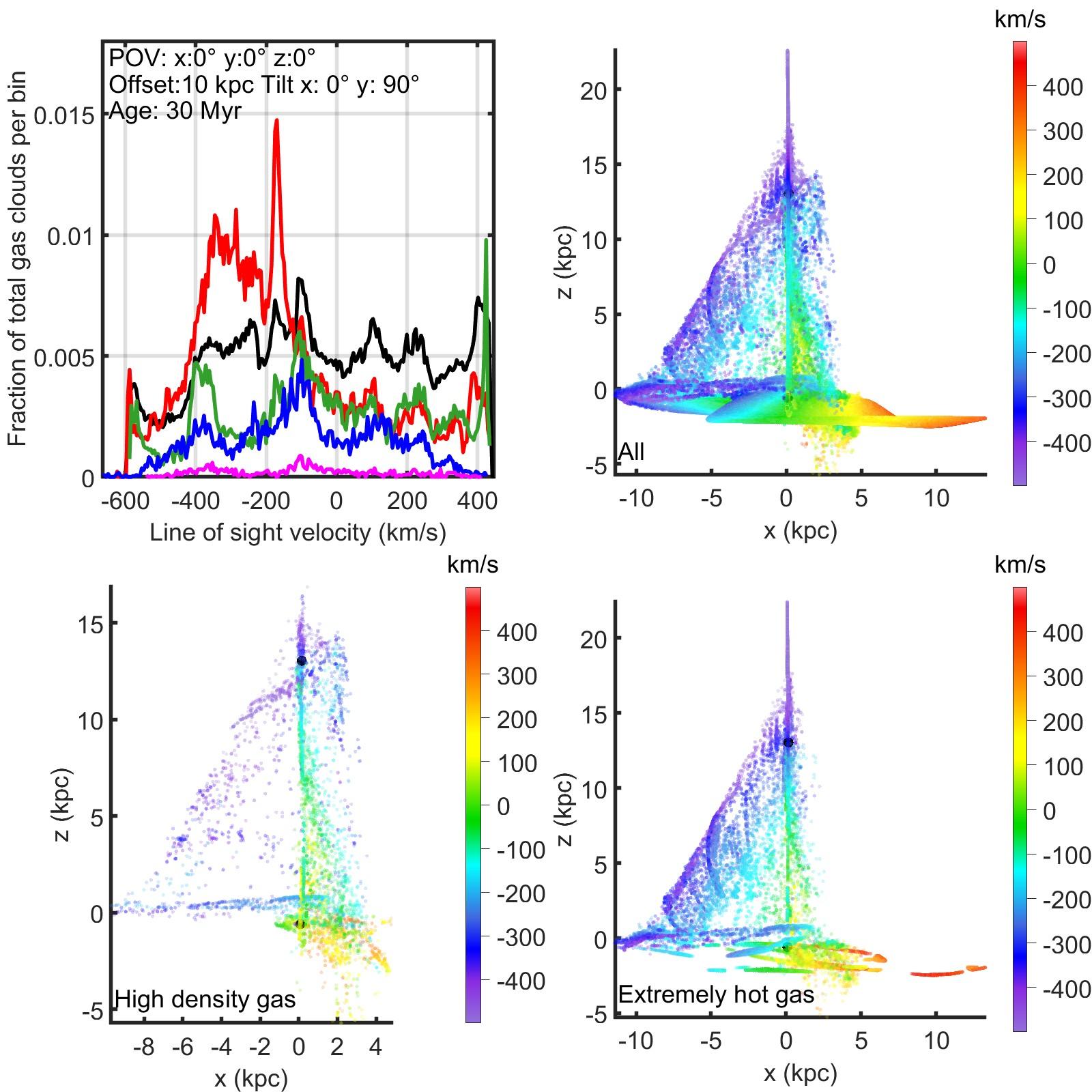}
   }
\caption{Edge-on view of G1. Age: 30 Myr. Collision: 90\degree{} inclination. Panel (a) 500 pc offset. Panel (b) The ten kiloparsec pc offset. Information on each sub-panel can be found in section \ref{sec:plot_description}.} %
\label{fig:90inc-rotx0y0z0}
 \end{center}
\end{figure}

At a maximum relative disk inclination of 90\degree{}, we can still see some of the effects of disk offset on the total particle spectra. \Cref{fig:90inc-rotx0y0z0} shows relatively flat spectra with peaks at -400, -200, 200, and 400 \kms{} for the 500 pc offset in panel (a). Compare this to the ten kiloparsec offset in panel (b) where the -400 \kms{} peak is reduced compared to the -200 \kms{} peak as now much more gas is found at velocities -400 \kms{} to -600 \kms. The individual cold gas components do not reflect these trends in the total and the previous cases. From this particular point of view, hot gas has more distinctive differences between the two offsets. It is primarily found at velocities of somewhat less than 0 \kms{} due to the counter-rotation of the galaxies. In the other panels, there are streams of gas visible moving to the left from the vertical G2 disk. 

\begin{figure}
  \begin{center}
   \subfigure[]{%
   \includegraphics[width=.5\textwidth]{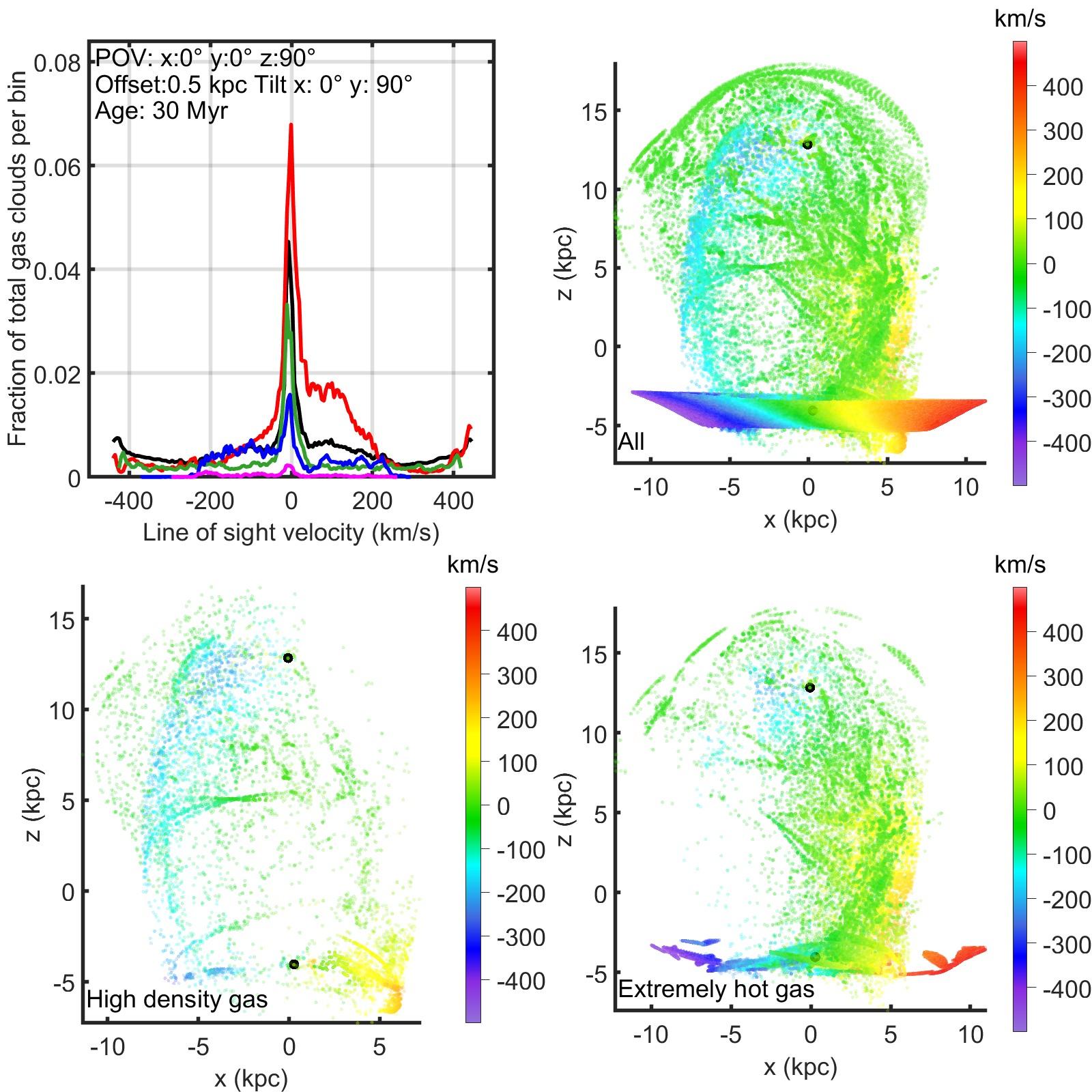}
   }
   \subfigure[]{%
   \includegraphics[width=.5\textwidth]{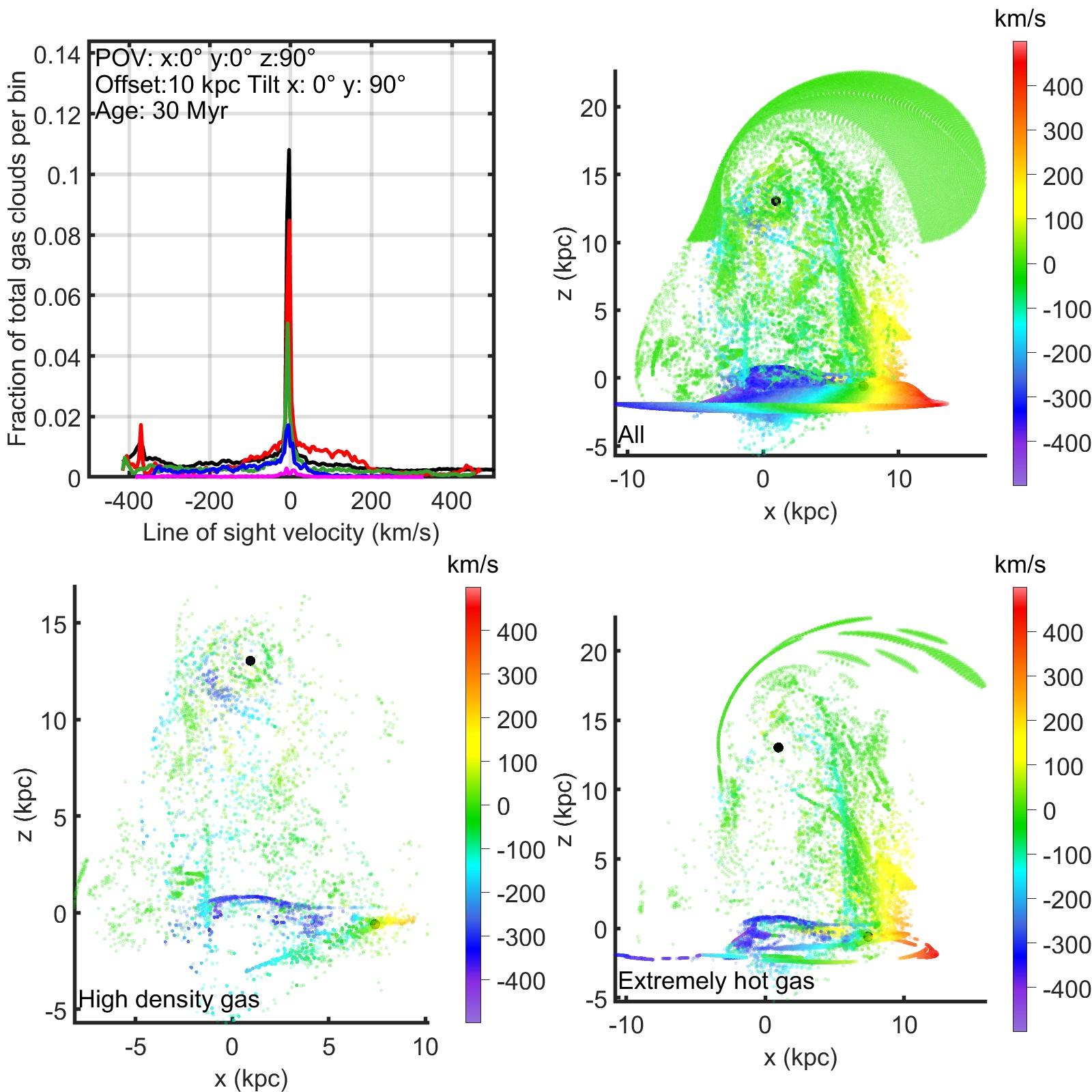}
   }
\caption{Edge-on view of G1, 90\degree{} relative to \Cref{fig:90inc-rotx0y0z0}. Age: 30 Myr. Collision: 90\degree{} inclination. Panel (a) 500 pc offset. Panel (b) The ten kiloparsec pc offset. Information on each sub-panel can be found in section \ref{sec:plot_description}.} %
\label{fig:90inc-rotx0y0z90}
 \end{center}
\end{figure}

In a low offset -90\degree{} inclination collision, the rotation of G1 and G2 produces streams of gas both to the right and left of G2, but the high offset means only one direction of gas flow is produced as the galaxies do not overlap enough for the rotation to create flows in two directions. This effect is apparent when comparing the \textit{extremely hot} gas spectra of panels (a) and (b) in \Cref{fig:90inc-rotx0y0z0}. In panel (b), the hot gas covers velocities from about -400 \kms to -200 \kms, but in the 500 pc offset collision in (a) the hot gas, velocities extend over a range from -400 \kms to 400 \kms. In the view where both G1 and G2 are seen edge-on, there is a small wing of velocities of 0 \kms{} to 100 \kms{} in the 500 pc offset collision, whereas the hot gas is seen only in a tight peak for the ten kiloparsec offset collision, see \Cref{fig:90inc-rotx0y0z90}.


Recall that in the spectra figures, the black line represents all phases of gas, red represents only \textit{extremely hot} gas ($T > $ \SI{e6}), blue represents \textit{high density} gas, and green represents gas initialized as \hh. These phases are defined in Section \ref{sec:spectra}. These spectra tend to consist of a single prominent peak with wings that span several hundred \kms. The physics driving the spectra toward a peak at 0 is momentum conservation in the dissipative collisions. This is the defining difference between a splash bridge and a tidal arm bridge. Only in the direct (and counter-rotating) gas disk collisions is so much of the momentum of the gas exchanged between galaxies. The offset and inclination at impact affect the overall shape of the spectra, but determining the collision parameters from the spectra alone is nearly impossible, given a strong dependence on the observation direction. However, the ratio of the \textit{extremely hot}/\textit{high density} spectra does appear to separate the high and low inclination and low and high offset parameter collisions.

In low-inclination collisions, the \textit{high density} component is dominant over the \textit{extremely hot}. By 30 Myr, the ten-kiloparsec offset 0\degree{} inclination produces a very high fraction of \textit{high density} gas. This provides further evidence that reducing the initial shock wave strength (at the high offset) and the subsequent turbulent shock cascades can rapidly turn the bridge gas into cool, dense gas clouds. The cold gas in the ten-kiloparsec offset, 0\degree{} collision does lead to many more gravitationally unstable clouds and, presumably, star formation as discussed in \citet{Yeager_2020}. While there are bursts of star formation in the 500 pc offset, 0\degree{} inclination case, the ten-kiloparsec offset collision produces gravitationally unstable clouds at a moderate but very steady rate, which begins about 27 Myr after the disk-disk collision. The star formation then continues at a nearly linearly decreasing rate until about 87 Myr after the collision before the second wave of increased star formation begins and continues to the end of that simulation at 117 Myr.

The window of star formation in the 500 pc offset collision lasts only a few million years. In contrast, the 500 pc offset, 0\degree{} inclination collision shows a small burst of star formation at 4 Myr after the disk-disk collision and then no more until about 30 Myr after the collision. These results are shown in Figure 6 of \citet{Yeager_2020}.

Hot gas in the low-inclination collisions does not show a peak near 0 \kms{} as in the higher inclination collisions. The ten-kiloparsec offset low-inclination collision, in particular, shows a very low amount of hot gas. This is not surprising as \Cref{fig:0inc-rotx0y0z0} shows a very depleted bridge when viewing the \textit{extremely hot} gas phase. Hot gas becomes more abundant as the relative inclination between the two gas disks increases. The exact range that the line-of-sight velocities span depends on the point of view, but typically, the range that the wings of the peak span is on the order of 300-400 \kms.

\section{Summary and Conclusions}

In this paper, we provide a broad overview of the thermal states, kinematics, and cloud collision turbulence in model splash bridges. In particular, we search for dependencies on the collision parameters of relative disk-disk central offset and inclination. The primary conclusion of this work is that there is a large variety of outcomes in each of these areas. This is due not only to strong dependencies on the primary collision parameters but also to the fact that many other parameters play some role. These include viewing angles, initial rotation speeds, and ISM distributions within the gas disks. We have not explored a full range of these and other galaxy parameters. For definiteness, the model collisions presented here are based on a 'typical' double spiral arm Sc galaxy and the parameters of the well-known splash bridge system, UGC 1294/5, the Taffy Galaxies. The Taffy Galaxies are likely best modeled as a collision of moderate disk inclinations ($<30\degree$). On the other hand, the Arp 194 system may provide a good high-inclination splash bridge prototype.

One interesting set of results from the models, which highlights the sensitivity to collision parameters, is how some gas splashed into the bridge reaccretes onto the parent galaxies. For example, in all 500 pc offset low-inclination disk-disk collisions, there is an immediate peak after impact in the amount of gas splashed from both G1 and G2, which is promptly followed by a reaccretion phase. This reaccretion reduces the fraction of gas within the splash bridges by $20\%$ to $75\%$ in about 10 Myr. In the case of 500 pc offset high-inclination disk-disk collisions, this rapid peak in the bridge gas fraction immediately after the beginning of the disk-disk collision does not occur. Instead, with the prolonged collision, gas splashes more slowly out of the disks, and the gas fraction in the bridge grows to a peak at around 50 Myr after the start of the disk-disk collision. The peak in highly inclined disk collisions matches the timing of a second splash bridge gas fraction peak in the low-inclination collisions. This second peak results from gas that falls back through the disks again (to the sides opposite the bridge) with sufficient velocity to push out more gas remaining in the disks after the initial impact. This second splash of gas out of the disks occurs for all collision parameters considered.

The line-of-sight distributions and profiles of hot and cold gas shown above are distinct from one another for nearly all disk-disk collisions, independent of the point of view, with few exceptions. Significant differences in the thermal evolution, depending on the relative disk offset and inclination of the collision, are prominent in the results. Collisions of low inclination produce by far the highest fractions of high-density phase gas, with high offsets (10 kiloparsecs) producing yet more high-density phase gas than low-offset (500 pc) collisions at a system age of 30 Myr. Increasing the relative inclination of the disks results in a much longer-lived hot gas fraction. The hot gas is supported by the turbulence of repeated cloud-cloud collisions that continue long after the initial disk-disk collision.

Central bridge disks form near the middle of splash bridges produced from low-offset $0\degree{}$ to $20\degree{}$ relative inclination disk collisions. These were first identified in \citet{yeager19}. These disks are built of the initially neutral hydrogen phase, highly compressed by shock waves reaching densities of giant molecular clouds. They do not contain ionized hydrogen gas. The central bridge disk consists of both extremely hot and high-density gas phases in disparate areas but contains very little of the initially dense gas. The disk begins ring oscillations and loses its disk-like appearance within ten million years, though the ring persists for 30-40 Myr. This may make it extremely difficult to observe central bridge disks as they will have evolved into a ring before much of the dense hydrogen appears.

Splash bridges are an important aspect of many galaxy collisions and their subsequent merger evolution. Models suggest that direct collisions between gas-rich disk galaxies can sometimes remove up to $60\%$ of the gas in the system. The interstellar gas can experience extreme heating with high relative velocities between the disks. Much of the gas stripped in splash bridges is later accreted onto the inner regions of the galaxy disks and is available to fuel nuclear activity. Returning gas from splash bridges ultimately forms stars, and these stellar populations will have unique and traceable kinematics. To fully understand the evolution of these bridges, it is important to obtain observations and specific models for more systems.

Our models show that splash bridges persist for timescales of hundreds of millions of years. For galaxy disk-disk collisions gravitationally bound to one another, the secondary collision may occur a couple of billion years after the first. Splashed gas clouds that are not accreted back onto either disk galaxy within the first 100 Myr are unlikely to accrete until the galaxies reverse on their post-collision trajectories due to the nature of their velocities relative to the still fast-moving galaxy disks. On billion-year timescales, though gas clouds bound under their self-gravity may persist, non-self-gravitating clouds will diffuse from pc to kiloparsec scales and be incorporated into the circumgalactic medium before their disk galaxies return.

Splash bridges are likely less rare than currently thought, indicated by the fact that our simulation shows that splash bridges can persist for over one hundred million years. In broad terms, splash bridges should be at least as common as ring galaxies. Ring galaxies require two disk galaxies to collide down the rotation axis of one another with fairly low offsets and relative inclinations. Generally, we observe rings in gas-rich disks because the increased star formation behind the ring wave is fairly obvious. Likely, many ring galaxies also contain splash bridges extending between the collided gas-rich disks. However, many systems likely collide along high relative inclinations or significant disk offsets that do not produce obvious rings but will still splash gas from their disks.

---
\section*{Acknowledgements}
This work was performed under the auspices of the U.S. Department of Energy by Lawrence Livermore National Laboratory under Contract DE-AC52-07NA27344.

\section*{Data Availability}
The data underlying this article will be shared on reasonable request to the corresponding author.

\bibliography{main}{}

\begin{thebibliography}{}
\makeatletter
\relax
\def\mn@urlcharsother{\let\do\@makeother \do\$\do\&\do\#\do\^\do\_\do\%\do\~}
\def\mn@doi{\begingroup\mn@urlcharsother \@ifnextchar [ {\mn@doi@}
  {\mn@doi@[]}}
\def\mn@doi@[#1]#2{\def\@tempa{#1}\ifx\@tempa\@empty \href
  {http://dx.doi.org/#2} {doi:#2}\else \href {http://dx.doi.org/#2} {#1}\fi
  \endgroup}
\def\mn@eprint#1#2{\mn@eprint@#1:#2::\@nil}
\def\mn@eprint@arXiv#1{\href {http://arxiv.org/abs/#1} {{\tt arXiv:#1}}}
\def\mn@eprint@dblp#1{\href {http://dblp.uni-trier.de/rec/bibtex/#1.xml}
  {dblp:#1}}
\def\mn@eprint@#1:#2:#3:#4\@nil{\def\@tempa {#1}\def\@tempb {#2}\def\@tempc
  {#3}\ifx \@tempc \@empty \let \@tempc \@tempb \let \@tempb \@tempa \fi \ifx
  \@tempb \@empty \def\@tempb {arXiv}\fi \@ifundefined
  {mn@eprint@\@tempb}{\@tempb:\@tempc}{\expandafter \expandafter \csname
  mn@eprint@\@tempb\endcsname \expandafter{\@tempc}}}

\bibitem[\protect\citeauthoryear{{Appleton}, {Charmandaris}  \&
  {Struck}}{{Appleton} et~al.}{1996}]{VIIZW4661996ApJ...468..532A}
{Appleton} P.~N.,  {Charmandaris} V.,   {Struck} C.,  1996, \mn@doi [\apj]
  {10.1086/177712}, \href
  {https://ui.adsabs.harvard.edu/abs/1996ApJ...468..532A} {468, 532}

\bibitem[\protect\citeauthoryear{{Appleton}, {Peterson}, {Helou}, {Jarrett},
  {Cluver}, {Ogle}, {Guillard}  \& {Boulanger}}{{Appleton}
  et~al.}{2011}]{2011AAS...21733527A}
{Appleton} P.~N.,  {Peterson} B.~W.,  {Helou} G.,  {Jarrett} T.~H.,  {Cluver}
  M.,  {Ogle} P.,  {Guillard} P.,   {Boulanger} F.,  2011, in American
  Astronomical Society Meeting Abstracts \#217. p. 335.27

\bibitem[\protect\citeauthoryear{{Appleton} et~al.,}{{Appleton}
  et~al.}{2015}]{appleton15}
{Appleton} P.~N.,  et~al., 2015, \mn@doi [\apj] {10.1088/0004-637X/812/2/118},
  \href {https://ui.adsabs.harvard.edu/#abs/2015ApJ...812..118A} {812, 118}

\bibitem[\protect\citeauthoryear{{Appleton} et~al.,}{{Appleton}
  et~al.}{2019}]{almataffy2019}
{Appleton} P.~N.,  et~al., 2019, in American Astronomical Society Meeting
  Abstracts \#233. p. 368.05

\bibitem[\protect\citeauthoryear{Appleton et~al.,}{Appleton
  et~al.}{2022}]{Appleton_2022}
Appleton P.~N.,  et~al., 2022, \mn@doi [The Astrophysical Journal]
  {10.3847/1538-4357/ac63b2}, 931, 121

\bibitem[\protect\citeauthoryear{{Borne}, {Lucas}, {Appleton}, {Struck},
  {Schultz}  \& {Spight}}{{Borne}
  et~al.}{1994}]{hstcartwheel1994AAS...185.6806B}
{Borne} K.~D.,  {Lucas} R.,  {Appleton} P.,  {Struck} C.,  {Schultz} A.,
  {Spight} L.,  1994, in American Astronomical Society Meeting Abstracts. p.
  68.06

\bibitem[\protect\citeauthoryear{{Braine} et~al.,}{{Braine}
  et~al.}{2003}]{braine03}
{Braine} J.,  et~al., 2003, in SF2A-2003: Semaine de l'Astrophysique Francaise.
  p.~231

\bibitem[\protect\citeauthoryear{{Braine}, {Lisenfeld}, {Duc}, {Brinks},
  {Charmandaris}  \& {Leon}}{{Braine} et~al.}{2004}]{Braine2004}
{Braine} J.,  {Lisenfeld} U.,  {Duc} P.~A.,  {Brinks} E.,  {Charmandaris} V.,
  {Leon} S.,  2004, \mn@doi [\aap] {10.1051/0004-6361:20035732}, \href
  {https://ui.adsabs.harvard.edu/abs/2004A&A...418..419B} {418, 419}

\bibitem[\protect\citeauthoryear{{Bushouse}}{{Bushouse}}{1987}]{1987ApJ...320...49B}
{Bushouse} H.~A.,  1987, \mn@doi [\apj] {10.1086/165523}, \href
  {https://ui.adsabs.harvard.edu/abs/1987ApJ...320...49B} {320, 49}

\bibitem[\protect\citeauthoryear{{Condon}, {Helou}, {Sanders}  \&
  {Soifer}}{{Condon} et~al.}{1993}]{condon93}
{Condon} J.~J.,  {Helou} G.,  {Sanders} D.~B.,   {Soifer} B.~T.,  1993, \mn@doi
  [\aj] {10.1086/116549}, \href
  {https://ui.adsabs.harvard.edu/#abs/1993AJ....105.1730C} {105, 1730}

\bibitem[\protect\citeauthoryear{{Condon}, {Helou}  \& {Jarrett}}{{Condon}
  et~al.}{2002}]{2002AJ....123.1881C}
{Condon} J.~J.,  {Helou} G.,   {Jarrett} T.~H.,  2002, \mn@doi [\aj]
  {10.1086/339558}, \href
  {https://ui.adsabs.harvard.edu/abs/2002AJ....123.1881C} {123, 1881}

\bibitem[\protect\citeauthoryear{Das \& Gronke}{Das \&
  Gronke}{2024}]{das2024magnetic}
Das H.~K.,  Gronke M.,  2024, Monthly Notices of the Royal Astronomical
  Society, 527, 991

\bibitem[\protect\citeauthoryear{Draine}{Draine}{2011}]{drainephysics}
Draine B.~T.,  2011, The Physics of the Interstellar and Intergalactic Medium.
Princeton University Press, 41 Williams Street, Princeton University Press, 6
  Oxford Street, Woodstock, Oxfordshire OX20 1TW

\bibitem[\protect\citeauthoryear{EAGLE}{EAGLE}{2020}]{eagle}
EAGLE 2020, The Eagle Project, \url {https://icc.dur.ac.uk/Eagle/}

\bibitem[\protect\citeauthoryear{FIRE}{FIRE}{2022}]{fire}
FIRE 2022, FIRE: Feedback In Realistic Environments, \url
  {https://fire.northwestern.edu/}

\bibitem[\protect\citeauthoryear{{Fogarty} et~al.,}{{Fogarty}
  et~al.}{2011}]{fog11}
{Fogarty} L.,  et~al., 2011, \mn@doi [\mnras]
  {10.1111/j.1365-2966.2011.19066.x}, \href
  {https://ui.adsabs.harvard.edu/#abs/2011MNRAS.417..835F} {417, 835}

\bibitem[\protect\citeauthoryear{{Gao}}{{Gao}}{2005}]{2005xmm..prop..169G}
{Gao} Y.,  2005, {X-raying the ``Interaction Zone`` of ``Taffy'' Galaxies},
  XMM-Newton Proposal

\bibitem[\protect\citeauthoryear{{Gao}, {Zhu}  \& {Seaquist}}{{Gao}
  et~al.}{2003}]{gao03}
{Gao} Y.,  {Zhu} M.,   {Seaquist} E.~R.,  2003, \mn@doi [\aj] {10.1086/378611},
  \href {https://ui.adsabs.harvard.edu/#abs/2003AJ....126.2171G} {126, 2171}

\bibitem[\protect\citeauthoryear{{Gerber}, {Lamb}  \& {Balsara}}{{Gerber}
  et~al.}{1992}]{1992ApJ...399L..51G}
{Gerber} R.~A.,  {Lamb} S.~A.,   {Balsara} D.~S.,  1992, \mn@doi [\apjl]
  {10.1086/186604}, \href
  {https://ui.adsabs.harvard.edu/abs/1992ApJ...399L..51G} {399, L51}

\bibitem[\protect\citeauthoryear{{Guillard}, {Boulanger}, {Pineau Des
  For{\^e}ts}  \& {Appleton}}{{Guillard} et~al.}{2009}]{gui09}
{Guillard} P.,  {Boulanger} F.,  {Pineau Des For{\^e}ts} G.,   {Appleton}
  P.~N.,  2009, \mn@doi [\aap] {10.1051/0004-6361/200811263}, \href
  {https://ui.adsabs.harvard.edu/#abs/2009A&A...502..515G} {502, 515}

\bibitem[\protect\citeauthoryear{{Hernquist}}{{Hernquist}}{1990}]{hernquist1990}
{Hernquist} L.,  1990, \mn@doi [The Astrophysical Journal] {10.1086/168845},
  \href {https://ui.adsabs.harvard.edu/\#abs/1990ApJ...356..359H} {356, 359}

\bibitem[\protect\citeauthoryear{Higdon, Buta  \& Purcell}{Higdon
  et~al.}{1998}]{higdon1998optical}
Higdon J.~L.,  Buta R.~J.,   Purcell G.~B.,  1998, The Astronomical Journal,
  115, 80

\bibitem[\protect\citeauthoryear{{Hopkins}, {Kere{\v{s}}}, {O{\~n}orbe},
  {Faucher-Gigu{\`e}re}, {Quataert}, {Murray}  \& {Bullock}}{{Hopkins}
  et~al.}{2014}]{2014MNRAS.445..581H}
{Hopkins} P.~F.,  {Kere{\v{s}}} D.,  {O{\~n}orbe} J.,  {Faucher-Gigu{\`e}re}
  C.-A.,  {Quataert} E.,  {Murray} N.,   {Bullock} J.~S.,  2014, \mn@doi
  [\mnras] {10.1093/mnras/stu1738}, \href
  {https://ui.adsabs.harvard.edu/abs/2014MNRAS.445..581H} {445, 581}

\bibitem[\protect\citeauthoryear{{Horellou} \& {Combes}}{{Horellou} \&
  {Combes}}{2001}]{combesmodel2001Ap&SS.276.1141H}
{Horellou} C.,  {Combes} F.,  2001, \mn@doi [\apss] {10.1023/A:1017524632342},
  \href {https://ui.adsabs.harvard.edu/abs/2001Ap&SS.276.1141H} {276, 1141}

\bibitem[\protect\citeauthoryear{ILLUSTRIS}{ILLUSTRIS}{2018}]{illustris}
ILLUSTRIS 2018, The Illustris simulation, \url
  {https://www.illustris-project.org/}

\bibitem[\protect\citeauthoryear{{Jarrett}, {Helou}, {Van Buren}, {Valjavec}
  \& {Condon}}{{Jarrett} et~al.}{1999}]{1999AJ....118.2132J}
{Jarrett} T.~H.,  {Helou} G.,  {Van Buren} D.,  {Valjavec} E.,   {Condon}
  J.~J.,  1999, \mn@doi [\aj] {10.1086/301080}, \href
  {https://ui.adsabs.harvard.edu/abs/1999AJ....118.2132J} {118, 2132}

\bibitem[\protect\citeauthoryear{{Joshi} et~al.,}{{Joshi}
  et~al.}{2019}]{2019ApJ...878..161J}
{Joshi} B.~A.,  et~al., 2019, \mn@doi [\apj] {10.3847/1538-4357/ab2124}, \href
  {https://ui.adsabs.harvard.edu/abs/2019ApJ...878..161J} {878, 161}

\bibitem[\protect\citeauthoryear{{Komugi} et~al.,}{{Komugi}
  et~al.}{2012}]{2012ApJ...757..138K}
{Komugi} S.,  et~al., 2012, \mn@doi [\apj] {10.1088/0004-637X/757/2/138}, \href
  {https://ui.adsabs.harvard.edu/abs/2012ApJ...757..138K} {757, 138}

\bibitem[\protect\citeauthoryear{{Marziani}, {Dultzin-Hacyan}, {D'Onofrio}  \&
  {Sulentic}}{{Marziani} et~al.}{2003}]{arp194marziani2003}
{Marziani} P.,  {Dultzin-Hacyan} D.,  {D'Onofrio} M.,   {Sulentic} J.~W.,
  2003, \mn@doi [\aj] {10.1086/368142}, \href
  {https://ui.adsabs.harvard.edu/abs/2003AJ....125.1897M} {125, 1897}

\bibitem[\protect\citeauthoryear{McAlpine et~al.,}{McAlpine
  et~al.}{2016}]{mcalpine2016eagle}
McAlpine S.,  et~al., 2016, Astronomy and computing, 15, 72

\bibitem[\protect\citeauthoryear{{Nicastro} et~al.,}{{Nicastro}
  et~al.}{2018}]{Nicastro2018}
{Nicastro} F.,  et~al., 2018, \mn@doi [\nat] {10.1038/s41586-018-0204-1}, \href
  {https://ui.adsabs.harvard.edu/\#abs/2018Natur.558..406N} {558, 406}

\bibitem[\protect\citeauthoryear{{Peterson} et~al.,}{{Peterson}
  et~al.}{2012}]{peterson12}
{Peterson} B.~W.,  et~al., 2012, \mn@doi [\apj] {10.1088/0004-637X/751/1/11},
  \href {https://ui.adsabs.harvard.edu/#abs/2012ApJ...751...11P} {751, 11}

\bibitem[\protect\citeauthoryear{{Peterson} et~al.,}{{Peterson}
  et~al.}{2018}]{2018ApJ...855..141P}
{Peterson} B.~W.,  et~al., 2018, \mn@doi [\apj] {10.3847/1538-4357/aaac2c},
  \href {https://ui.adsabs.harvard.edu/abs/2018ApJ...855..141P} {855, 141}

\bibitem[\protect\citeauthoryear{Roediger et~al.,}{Roediger
  et~al.}{2015}]{roediger2015stripped}
Roediger E.,  et~al., 2015, The Astrophysical Journal, 806, 103

\bibitem[\protect\citeauthoryear{Roussel et~al.,}{Roussel
  et~al.}{2007}]{Roussel_2007}
Roussel H.,  et~al., 2007, \mn@doi [The Astrophysical Journal]
  {10.1086/521667}, 669, 959

\bibitem[\protect\citeauthoryear{Santistevan, Wetzel, El-Badry, Bland-Hawthorn,
  Boylan-Kolchin, Bailin, Faucher-Giguère  \& Benincasa}{Santistevan
  et~al.}{2020}]{10.1093/mnras/staa1923}
Santistevan I.~B.,  Wetzel A.,  El-Badry K.,  Bland-Hawthorn J.,
  Boylan-Kolchin M.,  Bailin J.,  Faucher-Giguère C.-A.,   Benincasa S.,
  2020, \mn@doi [Monthly Notices of the Royal Astronomical Society]
  {10.1093/mnras/staa1923}, 497, 747

\bibitem[\protect\citeauthoryear{{Smith}, {Flynn}, {Candlish}, {Fellhauer}  \&
  {Gibson}}{{Smith} et~al.}{2015}]{smith15}
{Smith} R.,  {Flynn} C.,  {Candlish} G.~N.,  {Fellhauer} M.,   {Gibson} B.~K.,
  2015, \mn@doi [\mnras] {10.1093/mnras/stv228}, \href
  {https://ui.adsabs.harvard.edu/#abs/2015MNRAS.448.2934S} {448, 2934}

\bibitem[\protect\citeauthoryear{Sparre et~al.,}{Sparre
  et~al.}{2015}]{10.1093/mnras/stu2713}
Sparre M.,  et~al., 2015, \mn@doi [Monthly Notices of the Royal Astronomical
  Society] {10.1093/mnras/stu2713}, 447, 3548

\bibitem[\protect\citeauthoryear{Sparre, Hayward, Feldmann, Faucher-Giguère,
  Muratov, Kereš  \& Hopkins}{Sparre et~al.}{2016}]{10.1093/mnras/stw3011}
Sparre M.,  Hayward C.~C.,  Feldmann R.,  Faucher-Giguère C.-A.,  Muratov
  A.~L.,  Kereš D.,   Hopkins P.~F.,  2016, \mn@doi [Monthly Notices of the
  Royal Astronomical Society] {10.1093/mnras/stw3011}, 466, 88

\bibitem[\protect\citeauthoryear{{Struck}}{{Struck}}{2010}]{2010MNRAS.403.1516S}
{Struck} C.,  2010, \mn@doi [\mnras] {10.1111/j.1365-2966.2009.16224.x}, \href
  {https://ui.adsabs.harvard.edu/abs/2010MNRAS.403.1516S} {403, 1516}

\bibitem[\protect\citeauthoryear{{Struck}, {Appleton}, {Borne}  \&
  {Lucas}}{{Struck} et~al.}{1996}]{hstcartwheel1996AJ....112.1868S}
{Struck} C.,  {Appleton} P.~N.,  {Borne} K.~D.,   {Lucas} R.~A.,  1996, \mn@doi
  [\aj] {10.1086/118148}, \href
  {https://ui.adsabs.harvard.edu/abs/1996AJ....112.1868S} {112, 1868}

\bibitem[\protect\citeauthoryear{Tumlinson, Peeples  \& Werk}{Tumlinson
  et~al.}{2017}]{tumlinson2017circumgalactic}
Tumlinson J.,  Peeples M.~S.,   Werk J.~K.,  2017, Annual Review of Astronomy
  and Astrophysics, 55, 389

\bibitem[\protect\citeauthoryear{Vogelsberger et~al.,}{Vogelsberger
  et~al.}{2014}]{vogelsberger2014introducing}
Vogelsberger M.,  et~al., 2014, Monthly Notices of the Royal Astronomical
  Society, 444, 1518

\bibitem[\protect\citeauthoryear{{Vollmer}, {Braine}  \& {Soida}}{{Vollmer}
  et~al.}{2012}]{vollmer12}
{Vollmer} B.,  {Braine} J.,   {Soida} M.,  2012, \mn@doi [\aap]
  {10.1051/0004-6361/201219668}, \href
  {https://ui.adsabs.harvard.edu/#abs/2012A&A...547A..39V} {547, A39}

\bibitem[\protect\citeauthoryear{{Vollmer}, {Braine}, {Mazzilli-Ciraulo}  \&
  {Schneider}}{{Vollmer} et~al.}{2021}]{vollmer2021}
{Vollmer} B.,  {Braine} J.,  {Mazzilli-Ciraulo} B.,   {Schneider} B.,  2021,
  arXiv e-prints, \href {https://ui.adsabs.harvard.edu/abs/2021arXiv210107092V}
  {p. arXiv:2101.07092}

\bibitem[\protect\citeauthoryear{{Vshivkov}, {Lazareva}, {Snytnikov}, {Kulikov}
   \& {Tutukov}}{{Vshivkov} et~al.}{2011}]{2011ApJS..194...47V}
{Vshivkov} V.~A.,  {Lazareva} G.~G.,  {Snytnikov} A.~V.,  {Kulikov} I.~M.,
  {Tutukov} A.~V.,  2011, \mn@doi [\apjs] {10.1088/0067-0049/194/2/47}, \href
  {https://ui.adsabs.harvard.edu/abs/2011ApJS..194...47V} {194, 47}

\bibitem[\protect\citeauthoryear{{Wallin} \& {Struck-Marcell}}{{Wallin} \&
  {Struck-Marcell}}{1994}]{1994sacredmushroomstruck}
{Wallin} J.~F.,  {Struck-Marcell} C.,  1994, \mn@doi [\apj] {10.1086/174672},
  \href {https://ui.adsabs.harvard.edu/abs/1994ApJ...433..631W} {433, 631}

\bibitem[\protect\citeauthoryear{Wang, Ferland, Lykins, Porter, van Hoof  \&
  Williams}{Wang et~al.}{2014}]{wang15}
Wang Y.,  Ferland G.~J.,  Lykins M.~L.,  Porter R.~L.,  van Hoof P. A.~M.,
  Williams R. J.~R.,  2014, \mn@doi [Monthly Notices of the Royal Astronomical
  Society] {10.1093/mnras/stu514}, 440, 3100

\bibitem[\protect\citeauthoryear{Yeager \& Struck}{Yeager \&
  Struck}{2019}]{yeager19}
Yeager T.~R.,  Struck C.,  2019, \mn@doi [Monthly Notices of the Royal
  Astronomical Society] {10.1093/mnras/stz916}, 486, 2660, (Paper I)

\bibitem[\protect\citeauthoryear{{Yeager} \& {Struck}}{{Yeager} \&
  {Struck}}{2020a}]{Yeager_2020MNRAS.492.4892Y}
{Yeager} T.~R.,  {Struck} C.,  2020a, \mn@doi [\mnras] {10.1093/mnras/staa121},
  \href {https://ui.adsabs.harvard.edu/abs/2020MNRAS.492.4892Y} {492, 4892,
  (Paper II)}

\bibitem[\protect\citeauthoryear{{Yeager} \& {Struck}}{{Yeager} \&
  {Struck}}{2020b}]{Yeager_2020}
{Yeager} T.~R.,  {Struck} C.,  2020b, \mn@doi [The Astrophysical Journal]
  {10.3847/1538-4357/abc82a}, 905, 118, (Paper III)

\bibitem[\protect\citeauthoryear{{Zhu}, {Gao}, {Seaquist}  \& {Dunne}}{{Zhu}
  et~al.}{2007}]{2007AJ....134..118Z}
{Zhu} M.,  {Gao} Y.,  {Seaquist} E.~R.,   {Dunne} L.,  2007, \mn@doi [\aj]
  {10.1086/517996}, \href
  {https://ui.adsabs.harvard.edu/abs/2007AJ....134..118Z} {134, 118}

\makeatother
\end{thebibliography}
\bibliographystyle{mnras}

\bsp  
\label{lastpage}
\end{document}